\pdfoutput=1
\documentclass[a4paper,fleqn,usenatbib]{mnras}
\usepackage{newtxtext,newtxmath}
\usepackage[T1]{fontenc}
\usepackage{ae,aecompl}

\usepackage{graphicx}	
\usepackage{amsmath}	
\usepackage{amssymb}	
\usepackage{ulem}
\usepackage[french]{babel}
\usepackage[utf8]{inputenc}
\usepackage{numprint}
\usepackage{xcolor}

\newcommand{\ex}{\mathbf{e}_{\rm x}}

\newcommand{\ez}{\mathbf{e}_{\rm z}}
\newcommand{\rlight}{r_{\rm L}}
\newcommand{\Rs}{R_{\rm s}}

\definecolor{green}{rgb}{0.1,0.6,0.2}

\graphicspath{{/home/petri/publications/preparation/Quentin/article/}}

\title[General-relativistic pulsar emission]{General-relativistic pulsar radio and high-energy emission}

\author[Q. Giraud \& J. P\'etri]{
Quentin Giraud,\thanks{E-mail: quentin.giraud@astro.unistra.fr} J\'er\^ome P\'etri
\\
Universit\'e de Strasbourg, CNRS, Observatoire astronomique de Strasbourg, UMR 7550, F-67000 Strasbourg, France.
}

\date{Accepted XXX. Received YYY; in original form ZZZ}

\pubyear{2019}

\begin{document}
\label{firstpage}
\pagerange{\pageref{firstpage}--\pageref{lastpage}}
\maketitle

\begin{abstract}
According to current pulsar emission models, photons are produced within their magnetosphere or inside the current sheet outside the light-cylinder. Radio emission is favoured in the vicinity of the polar caps whereas the high-energy counterpart is presumably enhanced in regions around the light-cylinder, magnetosphere or/and wind. However, gravitational impacts on light-curves and their spectral properties have only been sparsely touched. In this paper, we present a new method to simulate the influence of the neutron star gravitational field on its emission according to general relativity. We numerically compute photon trajectories assuming a background Schwarzschild metric, applying our method to neutron star radiation mechanisms, like thermal emission from hot spots and non-thermal magnetospheric emission by curvature radiation. We detail the general-relativistic impacts onto observations made by a distant observer. Sky maps are computed using the vacuum electromagnetic field of a general-relativistic rotating dipole. We compare Newtonian results to their general-relativistic counterpart. For magnetospheric emission, we show that, more importantly than the aberration and the curvature of the trajectory of the photons, the Shapiro time delay significantly affected the phase delay between radio and high-energy light curves although the characteristic pulse profile that defines pulsar emission is kept.
\end{abstract}


\section{Introduction}
\label{sec:intro}

In 1967, Jocelyn Bell had observed a radio source that display with extreme regularity a peak of emission every $1.337$~seconds. This radio source, called pulsar for pulsating star, will later be identified as a neutron star in \citet{1968Natur.217..709H}, the collapsed core of a giant star stabilized by neutron degeneracy pressure. A neutron star has typically a diameter of $20$~kilometers and weight $1.5$~times the mass of the Sun \citep{ozel_masses_2016}. By using simple arguments about angular momentum and the magnetic flux conservation, this stellar remnant will also have a very high rotation speed with periods between 1~ms and 1~s and one of the strongest known magnetic field in the universe about the quantum critical value of $\numprint{4.4e9}$~T. 

Neutron stars are surrounded by a plasma formed of electrons/positrons pairs produced by photo disintegration in a strong magnetic field at the surface of the star \citep{ruderman_theory_1975}. This plasma corotates with the neutron star, by the action of the electromagnetic field, up to a limit called the light cylinder where the plasma rotation speed equals the speed of light~$c$ \citep{goldreich_pulsar_1969} and denoted by
\begin{equation}
\label{eq:RL}
\rlight = \frac{c}{\Omega}
\end{equation}
where $\Omega$ is the neutron star rotation pulsation. 
Beyond that limit, the magnetic field lines are assumed to be open, i. e. instead of joining the two magnetic poles, they leave one pole to infinity. To explain pulsar's characteristic emission, several models of  neutron star magnetospheres have been developed. They require empty gaps allowing for the existence of an electric field parallel to the magnetic field line, responsible for particle acceleration and radiation. Among the most popular models are the polar cap \citep{ruderman_theory_1975}, the outer gap \cite{cheng_energetic_1986} and the slot gap \citep{arons_pair_1983, dyks_two-pole_2003} with possible extension to the striped wind \citep{kirk_pulsed_2002, petri_unified_2011}. Charged particles accelerated by this electrical field are responsible for the pulsar emission, generating high-energy and radio emission by inverse Compton scattering, synchrotron radiation or, what we focus in this article, by curvature radiation. The pulse, periodically detected by an observer, is simply a consequence of the stellar magnetic field geometry and rotation.

Pulsars, especially those accreting, can also present two hot spots, located at the magnetic north and south poles where matter falls onto the neutron star's surface, with a thermal emission mainly in the X-ray band (around 100~eV). Because of their size, neutron stars have an important compactness defined by the ratio $\Rs/R = 0.5$ with $R$ the neutron star radius and $\Rs$ the Schwarzschild radius defined by $\Rs = \frac{2 \, G \, M}{c^{2}}$ with $M$ the mass of the star and $G$ the gravitational constant. This extreme compactness causes non negligible relativistic effects affecting the electromagnetic field structure and propagation of photons emitted in the vicinity of the neutron star. To determine how those effects affect terrestrial observations of pulsars, we simulate the trajectory of photons in the gravitational field of a neutron star by using ray tracing techniques. These techniques are mostly divided into two methods. A direct integration of the equation of motion in the prescribed metric has been implemented in \cite{vincent_gyoto:_2011}. It allows for ray tracing in a generic metric not necessarily analytical. Another approach performs the integration analytically in Schwarzschild or Kerr metric, leading to elliptical integrals as found by \cite{rauch_optical_1994}. This second technique is less general than the previous one but we found it much more accurate and faster than solving second order differential equations derived from the geodesic equations. We also include the calculation of the time of flight of the photon (Shapiro delay), as in \cite{2007ApJ...670..668B}, to properly compute the non thermal magnetospheric emission and the thermal hot spot emission as received by an distant observer.

Electromagnetic activity around neutron stars is evidenced by its pulsed emission detected on space \citep{abdo_second_2013} and ground-based \citep{lyne_shape_1988} telescopes. More than 2000~pulsars are known today, each showing a unique distinctive fingerprint depicted by its pulse profile in radio, X-rays and gamma-rays. The multi-wavelength light-curve evolution offers a unique insight into the real nature of the emission mechanisms as well as on their location and spread within the magnetosphere. Pulsars are mainly known as radio emitters. Although they have been observed since the early days of the discovery of pulsars fifty years ago, radio pulsars did not furnish severe constraints on the magnetospheric geometry and emission physics. With the advent of Fermi/LAT, more than 250~pulsars are known to emit also pulsed gamma-rays \citep{abdo_second_2013}. Gamma-ray pulsars have sharpened our understanding of pulsar magnetospheres because contrary to radio pulsars, gamma-ray pulsars spend a substantial fraction of rotational kinetic energy into high energy radiation. The flux remains significant even above several GeV severely constraining the emission sites to be well above the polar cap in order to avoid too strong magnetic absorption in magnetic field close to the critical value of $4.4\times10^9$~T \citep{daugherty_gamma-ray_1996}.

Sharp features in the light curves are interpreted as caustic formation in the outer part of the magnetosphere due to the combined effect of aberration and retardation \citep{morini_inverse_1983, dyks_relativistic_2004}. Phase alignment between radio and gamma-ray pulses seen in some millisecond pulsars suggests that for these pulsars, radio and gamma-rays are produced at the same location, and according to \cite{venter_modeling_2012} corresponding to 30\% of the light cylinder radius. A comprehensive study of pulsar light-curve characterization was compiled by \cite{watters_atlas_2009} for the three main high energy models namely, polar cap, slot gap (two-pole caustic) and outer gap. See also \cite{venter_probing_2009} and later \cite{pierbattista_light-curve_2015, pierbattista_young_2016} for a similar investigation. Such atlas are useful to constrain the pulsar obliquity and the observer line of sight inclination as pulse profiles are very sensitive to these parameters. Some refinements to the previous traditional views where proposed like the inner core and annular gaps by \cite{qiao_inner_2004} with some observational signatures shown by \cite{qiao_annular_2007}. Others used altitude-limited outer and slot gaps or low altitude slot gap models to better fit the light-curves especially for millisecond pulsars \citep{abdo_discovery_2010, venter_modeling_2012}.

In this paper we self-consistently include general-relativistic effects such as light bending and Shapiro delay to compute pulsed radio and high-energy emission. We employ semi-analytical solutions for the electromagnetic field around a rotating dipole in a slowly rotating neutron star metric, generalizing the classical Deutsch solution \citep{deutsch_electromagnetic_1955} to realistic neutron stars treated as compact objects. We also re-explore the thermal radiation from hot spots on the neutron star surface. In Sec.~\ref{sec:model} we recall the magnetospheric and emission models, explaining also the photon trajectories integration techniques in Schwarzschild spacetime. Sec.~\ref{sec:test} is devoted to the check of our algorithm by computing single photon trajectories as well as some images of the neutron star surface as seen by a distant observer. The measured fluxes of thermal hot spots is then shown in Sec.~\ref{sec:flux}. Eventually, high energy as well as radio emission maps are investigated in depth in Sec.~\ref{sec:emission}. Conclusions are drawn in Sec.~\ref{sec:conclusions}.

\section{Emission model}
\label{sec:model}

The emission model used in our model has been thoroughly described in \cite{petri_general-relativistic_2018}. We do not reproduce here the full details of this model but just remind some important features and extensions compared to \cite{petri_general-relativistic_2018}.

In any model of neutron star magnetospheric emission, we require several ingredients, namely
\begin{enumerate}
\item an accurate description of possible emission sites according to the existing magnetic field. We use a rotating vacuum magnetic dipole in general relativity for which excellent numerical approximations have been computed by \cite{petri_multipolar_2017-1}. There it is demonstrated that frame-dragging is irrelevant therefore neglected in our subsequent study. The Schwarzschild metric prevails as the background gravitational field.
\item a dynamical description of radiating particles and their composition. They are constrained to follow magnetic field lines in the corotating frame, mainly subject to curvature radiation along the local direction of field lines.
\item non thermal radiation process resulting from particle motion in the electromagnetic field. Synchrotron, curvature and inverse Compton emission are possible mechanisms for high and very high-energy photons. Here we focus on curvature radiation but results easily apply to any radiation fields although the spectra and pulse profile could slightly differ from one to another.
\item thermal emission from the polar caps producing an isotropic emission pattern in X-rays.
\item light bending induced by the stellar gravitational field. It is taken into account to produce sky maps. Moreover, the contribution of the Shapiro delay is now also included to compute light-curves.
\end{enumerate}
These items are touched upon in the following paragraphs. We then end this section by a discussion of the numerical algorithm used to produce our pulsar light-curves.

\subsection{Electromagnetic topology}

Since the work of \citet{deutsch_electromagnetic_1955} an exact analytical expression for a rotating magnetic dipole rotating in vacuum is known. The general-relativistic extension to his solution was found by \citep{petri_multipolar_2017-1}, using a semi-analytical radial expansion into rational Chebyshev functions leading to generalized spherical Hankel functions for outgoing waves and denoted by~$\mathcal{H}_\ell^{(1)}$. When the metric tends to minkowski spacetime, they reduce to the standard spherical Hankel functions~$h_\ell^{(1)}$ \citep{arfken_mathematical_2005}.

Denoting the angle between the magnetic axis and the rotation axis by~$\chi$, the general-relativistic solution for the magnetic field is advantageously decomposed into an aligned component, with weight $\cos\chi$ and using spherical Boyer-Lindquist coordinates $(r,\vartheta,\varphi)$, such that it equals the solution given by \cite{ginzburg_gravitational_1964}
\begin{subequations}
	\label{eq:MagneticStatic}
	\begin{align}
	\label{eq:MagneticStaticR}
	B^{\hat r}_\parallel & = - 6 \, B \, R^3 \, \left[ {\rm ln} \left( 1 - \frac{R_s}{r} \right) + \frac{R_s}{r} + \frac{R_s^2}{2\,r^2} \right] \, \frac{\cos\vartheta}{R_s^3} \\
	\label{eq:MagneticStaticT}
	B^{\hat \vartheta}_\parallel & = 3 \, B \, R^3 \, \left[ 2 \, \sqrt{ 1 - \frac{R_s}{r}} \, {\rm ln} \left( 1 - \frac{R_s}{r} \right) + \frac{R_s}{r} \, \frac{2\,r-R_s}{\sqrt{r\,(r-R_s)}} \right] \, \frac{\sin\vartheta}{R_s^3} \\
	B^{\hat \varphi}_\parallel & = 0
	\end{align}
\end{subequations}
and for the perpendicular component, with weight $\sin\chi$, by
\begin{subequations}
	\begin{align}
	\label{eq:DeutschEM}
	B^{\hat r}_\perp(\mathbf{r},t) & = \sqrt{\frac{3}{\pi}} \, \frac{f^{\rm B}_{1,1}(R)}{2\,r} \, \frac{\mathcal{H}^{(1)}_1(k\,r)}{\mathcal{H}^{(1)}_1(k\,R)} \, \sin \vartheta \, e^{i\,\psi} \\
	B^{\hat \vartheta}_\perp(\mathbf{r},t) & = \sqrt{\frac{3}{\pi}} \, \frac{f^{\rm B}_{1,1}(R)}{4} \,\times \\
	& \left[ \frac{\alpha}{r} \, \frac{\frac{d}{dr} \left( r \, \mathcal{H}^{(1)}_1(k\,r) \right)}{\mathcal{H}^{(1)}_1(k\,R)} + \frac{\Omega^2 \, R}{\alpha \, \alpha_R^2\,c^2} \, \frac{\mathcal{H}^{(1)}_2(k\,r)}{\frac{d}{dr} \left( r \, \mathcal{H}^{(1)}_2(k\,r) \right) |_{R}} \right] \,  \cos \vartheta \, e^{i\,\psi} \\
	B^{\hat \varphi}_\perp(\mathbf{r},t) & = \sqrt{\frac{3}{\pi}} \, \frac{f^{\rm B}_{1,1}(R)}{4} \,\times \\
	&  \left[ \frac{\alpha}{r} \, \frac{\frac{d}{dr} \left( r \, \mathcal{H}^{(1)}_1(k\,r) \right)}{\mathcal{H}^{(1)}_1(k\,R)} + \frac{\Omega^2 \, R}{\alpha \, \alpha_R^2\,c^2} \, \frac{\mathcal{H}^{(1)}_2(k\,r)}{\frac{d}{dr} \left( r \, \mathcal{H}^{(1)}_2(k\,r) \right) |_{R}} \, \cos 2\vartheta \right] \, i \, \, e^{i\,\psi} .
	\end{align}
\end{subequations}
$B$ is the magnetic field strength at the equator and
\begin{subequations}
	\begin{align}
	\label{eq:C1}
	\alpha_R & = \sqrt{ 1 - \frac{R_s}{R} }\\
 f^{\rm B}_{1,0}(r) & = - 4 \, \sqrt{3\,\upi} \, \frac{B\,R^3}{\Rs^2} \, \left[ \frac{\ln \left(1-x\right)}{x} + 1 + \frac{x}{2} \right] \\
 f^{\rm B}_{1,1}(r) & = - \sqrt{2} \, f^{\rm B}_{1,0}(r) \\
 x & = \Rs/r
 	\end{align}
\end{subequations}
\cite{rezzolla_general_2001} and \cite{rezzolla_electromagnetic_2004} found similar expressions about general-relativistic rotating dipoles in vacuum without numerical integration. Note also that $c/\Omega$ is not equal to the light-cylinder radius $\rlight$ because $\Omega$ is not the actual rotation rate of the neutron star as seen by a local observer. Indeed, the light-cylinder radius in Schwarzschild spacetime~$\rlight$ is properly defined by the location where the corotation speed reaches the speed of light for a local observer with his own clock ticking with proper time $d\tau = \alpha \, dt$. There the speed of light is reached for $r\,\Omega=\alpha\,c$ leading to an approximate expression given by
\begin{equation}
\rlight \approx \frac{c}{\Omega} \, \left( 1 - \frac{1}{2} \, \frac{\Omega\,\Rs}{c} - \frac{3}{8} \, \frac{\Omega^2\,\Rs^2}{c^2} \right).
\end{equation}
We use this value for the light-cylinder radius in general relativity. Polar cap shapes and separatrix locations are computed according to this expression. The difference between $\rlight$ and $c/\Omega$ is prominent only for millisecond pulsars.

\subsection{Emission sites}

Although our approach can deal with any shape of emission regions, we focus on the two standard sites: polar caps and slot gaps.

\subsubsection{Polar cap}

Polar caps are supposed to efficiently produce radio photons. The altitude of emission, constrained by radio observations, ranges from several stellar radii up to a substantial fraction of the light-cylinder radius~$\rlight$, about 10\% of $\rlight$ \citep{mitra_comparing_2004, mitra_meterwavelength_2016}. In our prescription for emissivity, we distinguish between two cases. The first, thermal, emission pattern forces photons to be emitted isotropically outwards and not embracing the electromagnetic field topology. This should mimic the thermal X-ray radiation from the hot spots on the surface. The second, non-thermal emission rule forces photons to propagate tangentially to the particle motion in the corotating frame at their launching position. This second option represents the traditional view about coherent radio emission from pulsars.

\subsubsection{Slot gap}

High-energy emission must be put at higher altitude in order to circumvent strong magnetic photon absorption process in a too strong magnetic field \citep{erber_high-energy_1966}. A commonly used acceleration gap where radiation leaves the star is the slot gap \citep{arons_pair_1983}. It is a thin layer sticking on the last open field line surface, the so-called separatrix. Emission is maximal on this separatrix and decreases monotonically when moving out of this surface. In a simplified model, we assume this layer to be infinitely thin.

\subsection{Radiation properties and aberration}
\label{sec:aberration}

There are several ways for particles to produce photons. We could consider any radiation mechanism. However, for the investigation of light-curve shapes only, studying a generic emission process is enough. Looking for broadband spectra and phase-resolved polarization properties would certainly require to consider a specific high-energy emission process.

We assume an isotropic distribution of pitch angle for particles in the comoving frame, therefore the emissivity reproduces the same isotropic pattern. In order to get the emissivity in the inertial frame, we need to perform a Lorentz boost from the rest frame which does not necessarily coincide with the corotating frame, to the observer frame.  
The line of sight of the distant observer makes an angle~$\zeta$ with respect to the $z$ axis such that its direction points towards the unit vector
\begin{equation}
\mathbf{n}_{\rm obs} = \sin\zeta \, \ex + \cos\zeta \, \ez.
\end{equation}

By assumption, in several models, particles follow magnetic field lines in the corotating frame. Their distribution function is isotropic in the rest frame of the fluid. The aberration formula was originally used by \citet{dyks_two-pole_2003} switching from the corotating frame to the observer frame. It is given by the usual textbook expression between two observers moving with constant relative velocity~$\mathbf{v}$ with respect to each other. 

To construct light curves from the pulsar magnetospheric emission, we have to account for the fact what, when considering the photon trajectory, we leave the rotating pulsar frame of reference for a static frame of reference attached to the observer, resulting in the above mentioned aberration phenomenon. To simulate aberration effects properly, we replace $\mathbf{n}'$ the unit vector of the propagation's direction at the emission position of the photon in the pulsar rotating frame, by its counterpart $\mathbf{n}$ in the observer frame. 

Starting with the minkowskian metric, the components of the two unit vectors directed along the photon trajectory are affiliated by the Lorentz transformation given by
\begin{subequations}
	\begin{align}
	n'_{\parallel} = & \gamma \, (n_{\parallel}-\beta n^{0}) \\
	n'_{\perp} = & n_{\perp}
	\end{align}
\end{subequations}
where $n'_{\parallel}$ and $n_{\parallel}$ are the components of $\mathbf{n}'$ and $\mathbf{n}$ that are parallel to $\bbeta$, $\bbeta$ being the normalized velocity vector of the rotating frame which, in the case of the pulsar, is equal to $\bbeta = \dfrac{r \, \Omega}{c} \sin \theta \,  \mathbf{e}_\varphi$. $n'_{\perp}$ and $n_{\perp}$ are the components perpendicular to $\bbeta$ and $\gamma = \dfrac{1}{\sqrt{1-\beta^{2}}}$. With the Doppler factor $\eta$ defined by
\begin{equation}
\label{eq:eta}
\eta = \dfrac{1}{\gamma \, (1-\bbeta\cdot\mathbf{n})} = \gamma \, (1+\bbeta\cdot\mathbf{n}') ,
\end{equation}
we get the usual flat spacetime aberration formula such that
\begin{equation}
\mathbf{n} = \frac{1}{\eta} \left[ \mathbf{n}' + \gamma \left( \dfrac{\gamma}{\gamma+1} \, (\mathbf{\beta} \cdot \mathbf{n}') + 1 \right) \mathbf{\beta} \right]  .
\label{eq:aberration_restreinte}
\end{equation}
Note that these quantities are not equal to the Lorentz factor and velocity measured by a local observer when gravity is included, they are coordinate quantities not physical quantities. Indeed, in a general relativistic case, we can still use the aberration formula Eq.~(\ref{eq:aberration_restreinte}) to find $\mathbf{n}$ from $\mathbf{n}'$ but we need to substitute $\bbeta$ by $\bbeta_{\rm RG}$ and $\gamma$ by $\gamma_{\rm RG}$ such as 
\begin{subequations}
	\begin{align}
	\label{eq:betaRG}
	\mathbf{\beta}_{\rm RG} = & \dfrac{\mathbf{\beta}}{\sqrt{1-\frac{\Rs}{r}}} \\ 
	\gamma_{\rm RG} = & \dfrac{1}{\sqrt{1-\beta_{\rm RG} ^{2}}}.
\end{align}
\end{subequations}
These are indeed the velocity and Lorentz factor as measured by a local observer for whom the flat spacetime aberration formula is valid.

\subsection{Ray tracing in Schwarzschild metric}
\label{sec:simulations}

The radiating electromagnetic field used in this paper is extracted from semi-analytical general-relativistic expressions. Thus in order to keep our investigation self-consistent, photons have to be subject to bending, time delay and gravitational redshift. In the present study, we take into account the light bending and the Shapiro delay but do not consider spectral properties thus neglecting photon reddening. Moreover, frame dragging does not impact neither on the electromagnetic field nor on the photon trajectories. We therefore decided to keep only the Schwarzschild metric as a representative geometry around neutron stars. This approximation improves for slowly rotating pulsars with period higher than several tenths of milliseconds.

Ray tracing techniques around black holes have been developed by many authors in different contexts involving black holes or neutron stars \citep{vincent_gyoto:_2011, psaltis_ray-tracing_2012, chan_gray:_2013}. Basically two different approaches are used. The first one integrates the equations of motion starting from an initial position and with fixed constants of motion. This is usually easy to implement but becomes inaccurate for large distances and is computationally expensive. The second approach integrates analytically the trajectories that are then given as integrals to be computed by any quadrature method. The latter is generally faster and more accurate for large distances but more involved for arbitrary motion \citep{rauch_optical_1994} in a Kerr spacetime. It is also not applicable to a general metric, extracted for instance from dynamical spacetime simulations and for which equations integrals are not amenable to closed formulas. Nevertheless, as we have to integrate millions of photon paths we prefer the second quadrature technique which already proved its efficiency in computing pulsar light-curves \citep{petri_general-relativistic_2018}.

Frame dragging effects around neutron stars is not relevant for its electrodynamics, especially for the electromagnetic field induced by a rotating magnet in vacuum as shown by \cite{petri_general-relativistic_2018}. Rotation of spacetime can even be neglected for millisecond pulsars. We therefore assume that Schwarzschild spacetime faithfully depicts the gravitational field around pulsars, independently of their rotation rate. Thus we do not discuss the frame dragging phenomenon in this article. We focus on the Schwarzschild metric to describe the space-time geometry around a massive object like a neutron star. In spherical Boyer-Lindquist coordinates it is represented by only one free parameter, the Schwarzschild radius~$\Rs$,
\begin{equation}
ds^{2}=\left(1-\dfrac{\Rs}{r}\right)c^{2}dt^{2}-\left( 1-\dfrac{\Rs}{r}\right)^{-1}dr^{2}-r^{2}(\sin^{2}\theta d\phi^{2}+d\theta^{2})
\label{eq:metric}
\end{equation}
In this metric, the trajectory of a photon is always contained within a plane defined by the location of the mass~$M$ and the initial direction of propagation of that photon~$\mathbf{n}$. Therefore, we can use a two dimensional projection of the Schwarzschild metric, identifying the plane of the trajectory to the equatorial plane $\theta=\upi/2$ leading to
\begin{equation}
ds^{2} = \left(1 - \dfrac{\Rs}{r} \right)c^{2}dt^{2} - \left( 1-\dfrac{\Rs}{r}\right)^{-1} dr^{2} - r^{2} d\phi .
\label{eq:metric2D} 
\end{equation}
$r$ and $\phi$ are the polar coordinates of the photon in the plane of the trajectory. Within this induced metric, the coordinates distance~$r$ to the origin and angle~$\phi$, as used by \cite{1994ApJ...425..767G}, are related by the equation for the trajectory as
\begin{equation}
\phi(r) = \phi_{0} \pm \int_{r_{0}}^{r} \dfrac{bdr}{r^{2}\sqrt{1-\frac{b^{2}}{r^{2}}(1-\frac{\Rs}{r})}}
\label{eq:trajectory}
\end{equation}
with $\phi_{0}$ and $r_{0}$ the coordinates of the emission point, $b$ the impact parameter defined as
\begin{equation}
b = \dfrac{r_{0}}{\sqrt{1-\frac{\Rs}{r_{0}}}} \, \sin\theta
\label{eq:impact}
\end{equation}
and $\theta$ the angle between the radial direction and the photon emission direction~$\mathbf{n}$. The plus and minus sign in front of the integral applies to receding $dr/dt>0$ and approaching $dr/dt<0$ photons respectively. To be able to determine the position of the photon at infinity, we replace $r$ in equation~(\ref{eq:trajectory}) by the Binet transformation $u=\frac{1}{r}$, so when $r\rightarrow\infty$, we have $u = 0$ and the integral~(\ref{eq:trajectory}) becomes
\begin{equation}
\phi (u) = \phi_{0} \mp \int_{u_{0}}^{u} \frac{b \, du}{\sqrt{1-b^{2} \, u^{2}(1-\Rs \, u)}}
\label{eq:trajectory_u}
\end{equation}
with $u_{0} = \frac{1}{r_{0}}$. Note the reversal of sign in front of the integral with respect to eq.~(\ref{eq:trajectory}). More precisely equation~(\ref{eq:trajectory_u}) has a positive sign when the photon is falling on the origin of the gravitational field and a minus sign when it is leaving toward infinity (and inversely for equation~\ref{eq:trajectory}). 

If the impact parameter is inferior to a certain critical value given by $b_{\rm c} = 3\sqrt{3}/2$ \citep{1998reas.conf...66K}, the photon falls on the origin where the star is located. In certain cases, for $b > b_{\rm c}$, the photon approaches the star in a first stage, reach a reversal point and then recedes toward infinity. At this turning point, the radial coordinate of the photon trajectory reaches is minimum equal to $r_{\rm min}$. Formally, it is the root of the third order polynomial in~$u$ defined by $p(u) = 1 - b^{2} \, u^{2} \, (1-\Rs \, u)$.

Of course these equations only give the position of the photon in the plane adapted to the trajectory. For a general orientation of this plane in a full three-dimensional space, we apply 3~rotations to bring the photon's trajectory into this adapted frame by using the Euler angles. These rotations are recalled in Appendix~\ref{app:A}.

Computing the time coordinate of the photon follows the same line as for the trajectory. It is found with another integral given by \cite{1983ApJ...274..846P} and reads
\begin{equation}
t= t_{0}+\displaystyle \int_{r_{0}}^{r}  \dfrac{dr}{ \left(1-\frac{\Rs}{r} \right) \, \sqrt{1-\dfrac{b^{2}}{r^{2}} \, \left(1-\frac{\Rs}{r}\right)}}
\label{eq:temps}
\end{equation}
with $t_{0}$ the time coordinate of the date when the photon is emitted.
Once again we use the Binet substitution $u=\frac{1}{r}$ to rewrite it as
\begin{equation}
t = t_{0} - \displaystyle \int_{u_{0}}^{u} \dfrac{du}{u^{2}(1-\Rs \, u) \, \sqrt{1 - b^{2} \, u^{2} \, (1-\Rs \, u)}}
\label{eq:temps_u}
\end{equation}
All along this paper, we integrate the equations~(\ref{eq:trajectory_u}) and (\ref{eq:temps_u}) using the Clenshaw-Curtis quadrature explained in depth in \cite{press_numerical_2007}. It uses Fast Fourier Transform techniques employing cosines transforms to perform Chebyshev interpolation and integration.

\section{Test of photon trajectory integrations}
\label{sec:test}

Before applying the code to realistic pulsar magnetospheres, we test our integration scheme against simple cases such as photon trajectories in the equatorial plane and the image of a neutron star as perceived by a distant observer.

\subsection{Single photon motion in the equatorial plane}

\begin{figure}
	\includegraphics[width=\columnwidth]{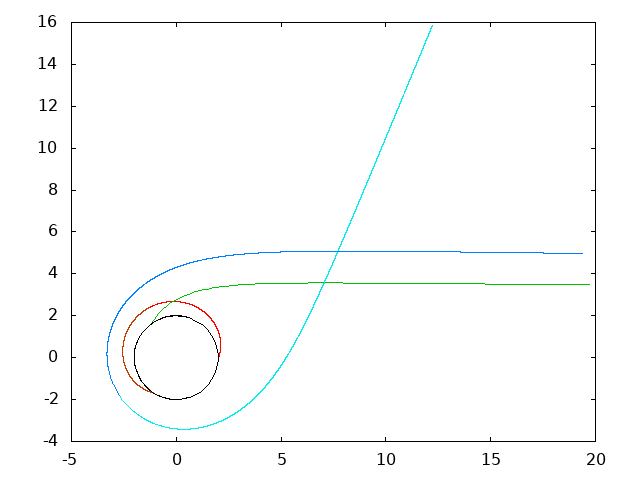}
	\caption{Some possible trajectories for a photon travelling around a Schwarzschild radius. Case (i) is shown green, case~(ii) in red and case~(iv) in blue.}
	\label{fig:traj}
\end{figure}

\begin{figure}
	\includegraphics[width=\columnwidth]{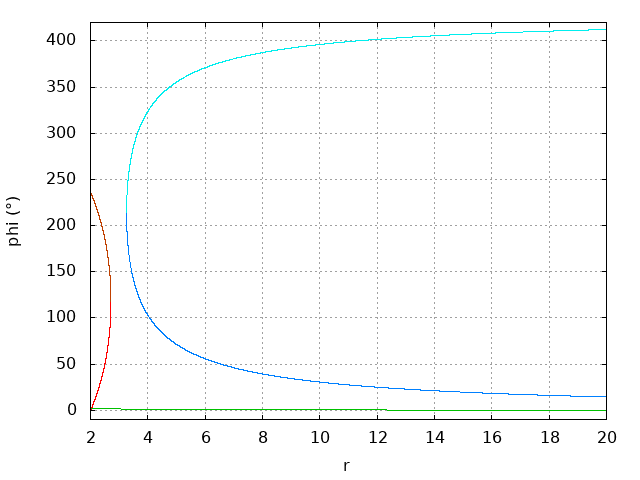}
	\caption{Evolution of the polar angular coordinate $\phi$ with respect to the radial coordinate~$r$ corresponding to the paths shown in Fig.~\ref{fig:traj}. The function$\phi(r)$ is double-valued for trajectories showing turning points.}
	\label{fig:phi_traj}
\end{figure}

Because of the spherical symmetry of the Schwarzschild metric, it is always possible to reduce the particle motion to a plane such that $\theta=\upi/2$. Photon trajectories around neutron stars are then of four kinds depending on their receding or approaching motion and depending on their capture by the horizon or not. We distinguish
\begin{enumerate}
\item photons produced at the surface and leaving the star, going to infinity.
\item photons produced at the surface and leaving the star but then returning to it.
\item photons coming from infinity and approaching the star to hit its surface.
\item photons coming from infinity and approaching the star, being deflected and then going back to infinity.
\end{enumerate}
Typical examples of the case (i), (ii) and (iv) are shown in Fig.~\ref{fig:traj}. Note that (iii) is similar to (i) except that the photon travels in the opposite direction. The Schwarzschild radius is normalized to $\Rs=2$ and depicted by the black circle. Case (i) in Fig.~\ref{fig:traj} shows a photon leaving the star from its surface. It has an impact parameter $b<b_{\rm c}$ and is therefore always captured by the horizon. This kind of trajectories is easily computed because the radius monotonically decreases with the polar angle $\phi$. This is seen by inspection of Fig.~\ref{fig:phi_traj} showing a monovalued function~$\phi(r)$.
Deflection of light in the vicinity of a compact object must be handled more carefully because the photon first approaches the star with decreasing radius~$r$ but at the turning point, it recedes to infinity by increasing again its radial coordinate~$r$. Such a trajectory is seen in Fig.~\ref{fig:traj}. The approaching part is colored in light blue and the receding part colored in dark blue. Therefore the function~$\phi(r)$ is multivalued and must be treated appropriately by cutting it into two parts separated by the minimal distance to the centre of the star~$r_{\rm min}$ as shown in Fig.~\ref{fig:phi_traj}. The integration constants in the integral formulation of the trajectory are chosen to smoothly join both parts of the motion at the reversal point. The minimal distance~$r_{\rm min}$ is found by analytically solving a third order polynomial. A last check has been performed for trajectories not expected in neutron stars but useful for black holes. In Fig.~\ref{fig:traj}, a photon coming out from the horizon is strongly deflected and then returns inside the horizon, red curve. Here also, the trajectory shows a turning point associated not to a minimal distance but to a maximal distance~$r_{\rm max}$ which is also found by solving a third order polynomial. The polar angle function~$\phi(r)$ is again multi-valued and must be separated in receding and returning parts as shown in Fig.~\ref{fig:phi_traj} in red line. Care must be taken to smoothly join both parts of the trajectory at the turning point corresponding to~$r_{\rm max}$.

\subsection{Image distortion of a neutrons star}
\label{sec:image}

To show how the gravitational field of a neutron star affects the trajectory of photons, we simulated the image seen by a distant observer of the neutron star surface. To do this we spread emission points all around its surface, located at twice the Schwarzschild radius $R=2\,\Rs$ thus a compactness of a neutron star typically of $K=\Rs/R = 0.5$), each point is separated from its neighbour by a difference of several degrees both in colatitude $\Delta \theta$ and in longitude~$\Delta\phi$. The image of this surface is obtained, for flat space-time associated to the the minkowski metric simply by tracing a line toward an hypothetical screen, a plane perpendicular to the line of sight of a distant observer. 

In the general-relativistic case, we search, with a root-finding function, the angle of emission $\theta$, in the interval $]-\frac{\upi}{2},\frac{\upi}{2}[$, for which the angle $\phi(r)$ given by equation~(\ref{eq:trajectory_u}) is the same as the position of the observer at infinity (here we take $\phi_{\rm obs}=0$ as the position of the observer), then we compute the impact parameter~$b$. This impact parameter~$b$ being also the distance between the photon's trajectory and a parallel line that comes from the origin when general-relativistic effects are negligible \citep{1998reas.conf...66K} so, at large distance $r\gg \Rs$, the impact point of the photon on the screen is at a distance~$b$ from the projection of the centre of the star on the line that is the intersection of the screen and the plane containing the trajectory of the photon.

\begin{figure}
	\includegraphics[width=\columnwidth]{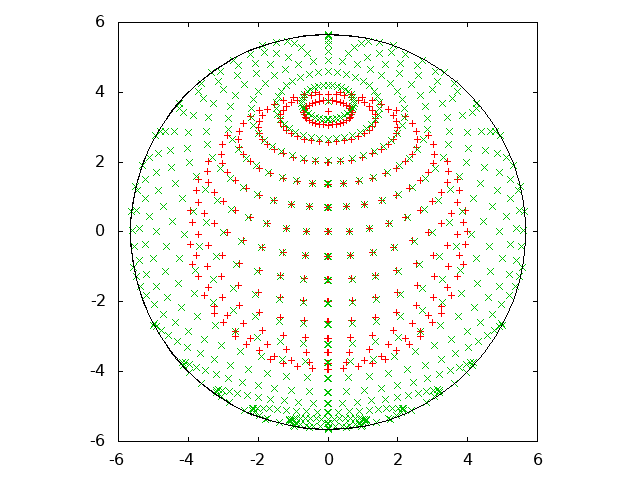}
    \caption{Image of the surface of the neutron star for a flat space time (in red) and with general-relativistic effects (in green) for a compactness of $K=0.5$ and an observer located in a direction of $\zeta=30\degr$. The black circle of radius $R_\infty$ is also shown for reference.}
    \label{fig:map30-re4}
\end{figure}

\begin{figure}
	\includegraphics[width=\columnwidth]{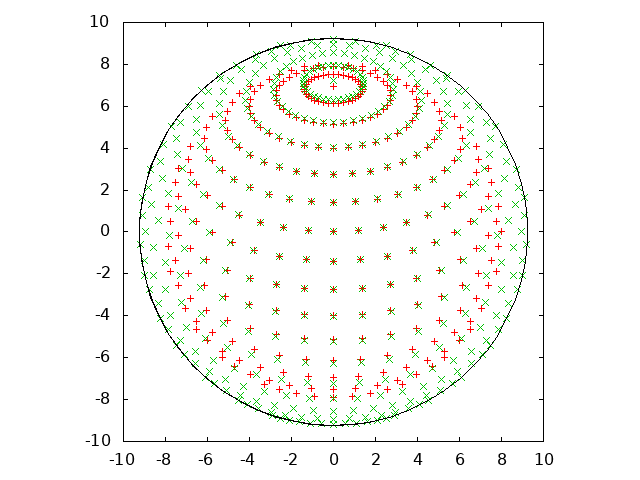}
    \caption{Image of the surface of the neutron star for a flat space time (in red) and with general-relativistic effects (in green) for a compactness of $K=0.25$ and an observer  located in a direction of $\zeta=30\degr$. The black circle of radius $R_\infty$ is also shown for reference.}
    \label{fig:map30-re8}
\end{figure}

We compare minkowskian to general-relativistic images for several line of sight inclination angles $\zeta$ and report the case $\zeta=30\degr$ in Fig.~\ref{fig:map30-re4}.
We observe that general-relativistic effects enlarge the image seen by a distant observer. The important point is that a larger portion of the stellar surface is visible because of light bending by the gravitational field of the star. This allows photons emitted from regions behind the star and thus normally hidden in the minkowskian metric to reach the observer. Those effects disappear progressively when the compactness decrease as we can see in Fig~\ref{fig:map30-re8} where the same image is found but for a compactness $K=0.25$ (i. e. a radius of the star that is equal to $8$ times its gravitational radius). 
The apparent radius of the neutron star, noted~$R_{\infty}$, is actually deduced from the impact parameter. Knowing that photons can only leave the star if $|\alpha|<\upi/2$, the apparent radius measured at spatial infinity immediately follows as
\begin{equation}
 R_{\infty} = \frac{R}{\sqrt{1-\frac{\Rs}{R}}} .
\end{equation}
$R$ is the radial coordinate labelling the boundary of the neutron star. High compactnesses imply large apparent radii $R_{\infty}$ having important implications for the measured flux, temperature and hot spot area.

\section{Thermal flux from the polar caps}
\label{sec:flux}

A first interesting application of ray tracing around neutron stars concerns its thermal X-ray emission from rotating hot spots located around the magnetic poles on the stellar surface. This emission is mostly seen in X-rays and useful to constrain the stellar mass-radius ratio $M/R$ therefore its compactness. In this section, we compute sky maps of X-ray light-curves similar to the sky maps employed for pulsed high-energy magnetospheric emission. We take full account of general-relativistic effects in a Schwarzschild spacetime, namely: light bending, redshift and Shapiro delay.

The polar caps, that is the regions delimited by the last closed field lines when they cross the neutron star's surface, are thought to be hot spots emitting like a black body of temperature around hundred eV therefore mainly observed in X-rays. The flux received by a distant observer from these two hotspots, assuming a pure dipole, is affected by general-relativistic effects induced by the mass of the neutron star. We simulate the flux emitted by the polar caps. We compare the results for a distant observer when residing in a flat space-time and in the Schwarzchild metric. 

Following the notations of \cite{2007ApJ...670..668B}, we introduce several angles such as the angle~$\alpha$ between the rotation axis and the magnetic axis, the angle~$\xi$ between the hot spot velocity vector and the direction of the line of sight expressed by
\begin{equation}\label{eq:cosxi}
\cos\xi = \dfrac{\sin\theta}{\sin\psi} \, \sin i \, \sin \varphi
\end{equation}
and the angle~$i$ between the rotation axis and the direction of the line of sight. Lastly, $\varphi$ is the pulsar's phase. The expression of the observed flux per unit frequency~$\nu$ emitted by each polar cap then reads
\begin{equation}
F(\nu) = \left( 1-\frac{\Rs}{R} \right)^{1/2} \, \eta^{4} \, I(\theta) \, \cos\theta \, \dfrac{\partial \cos\theta}{\partial\cos\psi}\frac{dS}{D^{2}}
\label{eq:Flux}
\end{equation}
where $I(\theta)$ is the intensity of the emission from one polar cap of surface area~$dS$. For the remainder of this section, we admit an isotropic emission pattern with a constant intensity~$I(\theta)$ not depending on $\theta$. $\eta$ is the Doppler factor measured by a local observer and expressed as 
\begin{equation}
\label{eq:Doppler}
\eta = \frac{1}{\gamma \, (1-\frac{\upsilon}{c} \cos\xi)}
\end{equation}
with its local 3-velocity 
\begin{equation}\label{eq:upsilon}
  \upsilon=\frac{2\,\upi\, R}{P \, \sqrt{1-\frac{\Rs}{R}}} \, \sin\alpha
\end{equation}
where $P$ is the pulsar rotation period. The Lorentz factor is simply related to this 3-velocity by
\begin{equation}\label{eq:gamma}
\gamma = \frac{1}{\sqrt{1-\frac{\upsilon^{2}}{c^{2}}}} .
\end{equation}
Considering $\psi$ as the polar cap position i.e. the angle between the magnetic axis and the line of sight \citep{2004AA...426..985V,2003MNRAS.343.1301P} we found
\begin{equation}
 \cos\psi = \pm \, (\cos i \cos\alpha + \sin i \sin\alpha \cos\varphi) .
\label{eq:psi}
\end{equation}
The plus sign corresponds to the primary polar cap whereas the minus sign to the antipodal polar cap (or secondary pole). $\psi$ is equal to $\phi(u)$ in equation~(\ref{eq:trajectory_u}) when $\phi_{0}=0$ and $r\rightarrow\infty$ (i.e. when $u$ is null) so we can find $\theta$ 
from equations~(\ref{eq:trajectory_u}) and (\ref{eq:psi}). We compute the received flux by using equation~(\ref{eq:Flux}). Thus
 \begin{equation}
F(\nu) = \sqrt{ 1 - \frac{\Rs}{R}} \, \eta^{4} \, I \, \cos\theta \, \dfrac{\sin\theta}{\sin\psi} \,\dfrac{\partial\theta}{\partial\psi} \, \dfrac{dS}{D^{2}} .
\label{eq:Flux2}
\end{equation}
With equation~(\ref{eq:trajectory}), we compute $\frac{\partial\psi}{\partial\theta}$ and obtain
\begin{equation}
\dfrac{\partial\psi}{\partial\theta} = \int_{R}^{\infty} \frac{b'(\theta) \,dr}{r^{2} \left[ 1 - \dfrac{b(\theta)^{2}}{r^{2}} \, \left(1-\frac{\Rs }{r}\right) \right]^{3/2}}
\label{eq:Dphi}
\end{equation}
or by the usual change of variable with $u=1/r$
\begin{equation}
\dfrac{\partial\psi}{\partial\theta} 
= \int_{0}^{u} \frac{b'(\theta)\,du}{\left[1-b(\theta)^{2} \, u^{2} \, (1-u \,\Rs )\right]^{3/2}}
\label{eq:Dphi_u}
\end{equation}
with the impact parameter derivative given by
\begin{equation}
b'(\theta)=\dfrac{\partial b}{\partial\theta}= \dfrac{R}{\sqrt{1-\frac{\Rs }{R}}} \cos\theta
\label{eq:Dimpact} .
\end{equation}
When $\psi\rightarrow 0$, we use the asymptotic limit $ \frac{\sin\theta}{\sin\psi} = \sqrt{1-\frac{\Rs }{R}}$
such that the flux simplifies into
\begin{equation}
F(\nu) = \left( 1 - \frac{\Rs}{R} \right) \, \eta^{4} \, I \, \cos\theta \, \dfrac{\partial\theta}{\partial\psi} \, \dfrac{dS}{D^{2}} .
\label{eq:Flux_asymptote}
\end{equation}
In the minkowskian flat space-time, there is no light bending effect i.e. $\cos \psi = \cos \theta$ and the received flux reduces to
\begin{equation}
F(\nu) = \sqrt{1-\frac{\Rs}{R}} \, \eta^{4} \, I \, \cos\theta \, \dfrac{dS}{D^{2}} .
\label{eq:Flux_approx}
\end{equation}
In all cases, the received flux must be considered as null if $\theta$ is not in the interval $[-\frac{\pi}{2};\frac{\pi}{2}]$ as photons cannot travel through the star (the other cases whenever $\theta \not\in [-\frac{\pi}{2};\frac{\pi}{2}]$ correspond to photons pointing towards the centre of the star through the crust and must be discarded).

The neutron star flux as measured on Earth is the sum of the flux emitted from both the polar caps. However, we need to add a delay to the actual phase in order to take into account the time of flight of photon in the Schwarzschild metric. In Minkowski space-time it is simply the time delay produced by the distance between the centre of star and the observer divided by the speed of light~$c$ plus the retarded time given by Roemer delay due to finite propagation speed of light
\begin{equation}\label{eq:tret}
t_{\rm ret} = - \frac{\mathbf{n}_{\rm obs} \cdot \mathbf{r}}{c}
\end{equation}
with $\mathbf{n}_{\rm obs}$ the unit vector directed toward the observer starting from the emission point.
However, in general relativity, the time of flight must be modifier to include the Shapiro delay following eq.(\ref{eq:temps_u}). 

Fully self-consistent and general-relativistic light curve computations require light bending, gravitational redshift and Shapiro delay. All these effects are now presented in several sky maps. In all situations, the observer is placed at large distances where gravitational effects can be neglected. Typically we set its distance to $D=10^4\,R$ where general-relativistic effects caused by the gravitational field of the neutron star are expected to remain less than $10^{-3}$. The neutron star obliquity is set to $\chi=45\degres$.

\begin{figure}
	\includegraphics[width=\columnwidth]{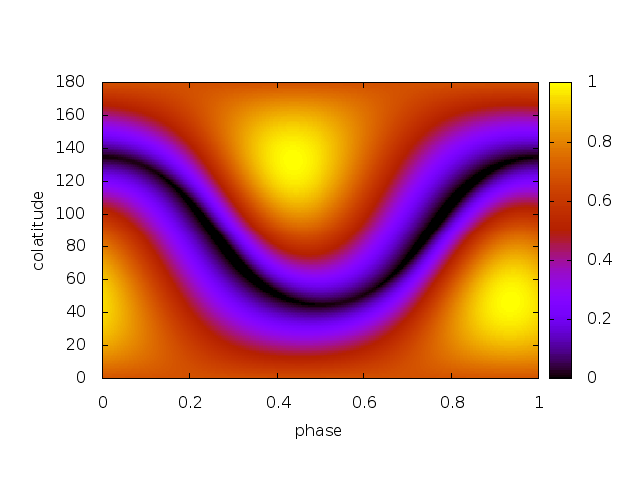} 
	\caption{Flux received from the two polar caps by a distant observer with an angle of $chi = 45\degr$ between the magnetic axis and the rotation axis in the minkowskian case.}
	\label{fig:Flux45_mink}
	\end{figure}

\begin{figure}
	\includegraphics[width=\columnwidth]{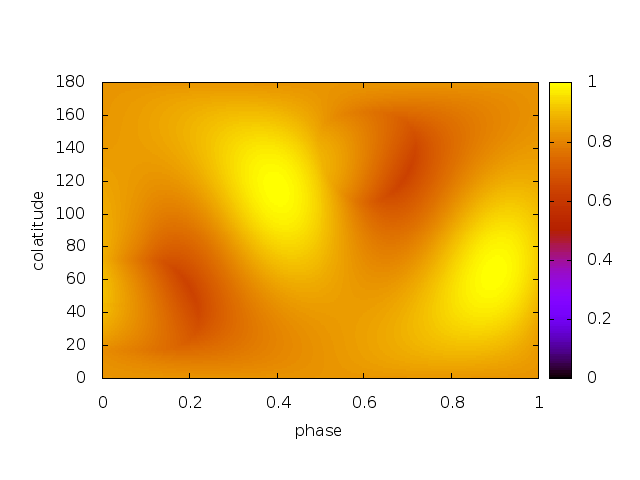}
	\caption{Flux received from the two polar caps by a distant observer with an angle of $\chi = 45\degr$ between the magnetic axis and the rotation axis in the relativistic case.}
	\label{fig:Flux45_rel}
\end{figure}

Fig.~\ref{fig:Flux45_mink} shows the flux received for the minkowskian metric where all the relativistic effects have been removed. A full period is normalized to phase equal to one an the maximum flux is also normalised. A S-shape black stripe with vanishing flux clearly separates both hot spots in the diagram. The two emission regions are well separated in the phase-inclination of line of sight plane. In general relativity, the situation is much less clear-cut as seen in the sky maps of Fig.~\ref{fig:Flux45_rel} representing the flux received from a neutron star of compactness $\Rs/R=0.5$. Both hot spots are visible a much larger fraction of the period with significant overlapping emission.

\begin{figure}
	\includegraphics[width=\columnwidth]{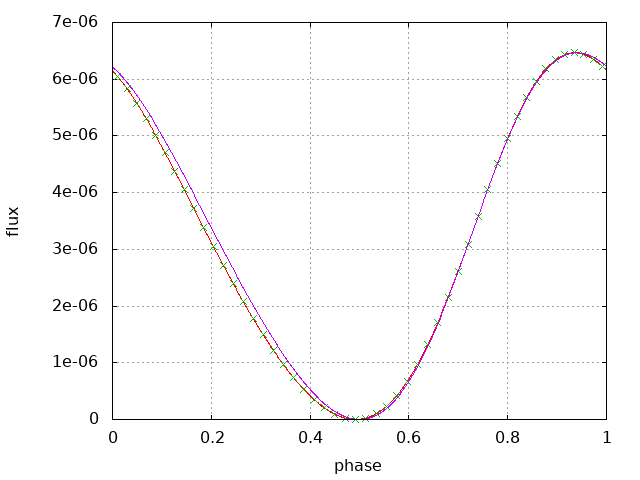}
	\caption{Flux received in the minkowskian case for a line of sight and a magnetic axis that form an angle of $45\degr$ with the rotation axis with in dotted line the emission from each of the polar cap, in red the sum of the two and in purple the sum without flight time. 
	}
	\label{fig:coupeF45-45M}
\end{figure}

\begin{figure}
	\includegraphics[width=\columnwidth]{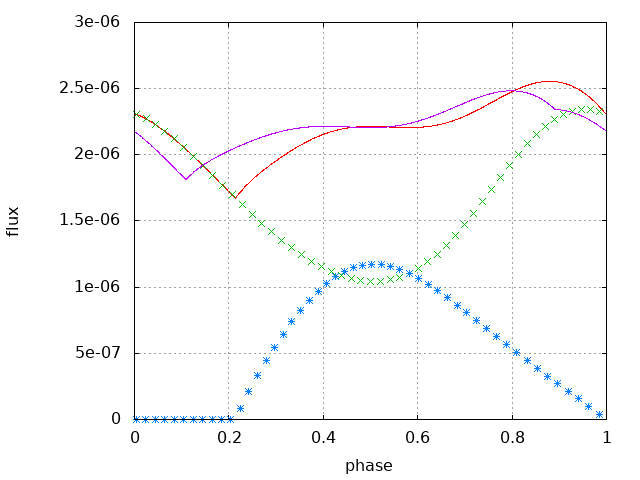} 
	\caption{Flux received in the relativistic case for a line of sight and a magnetic axis that form an angle of $45\degr$ with the rotation axis with in dotted line the emission from each of the polar cap, in red the sum of the two and in purple the sum without flight time.
	}
	\label{fig:coupeF45-45R}
\end{figure}

\begin{figure}
\includegraphics[width=\columnwidth]{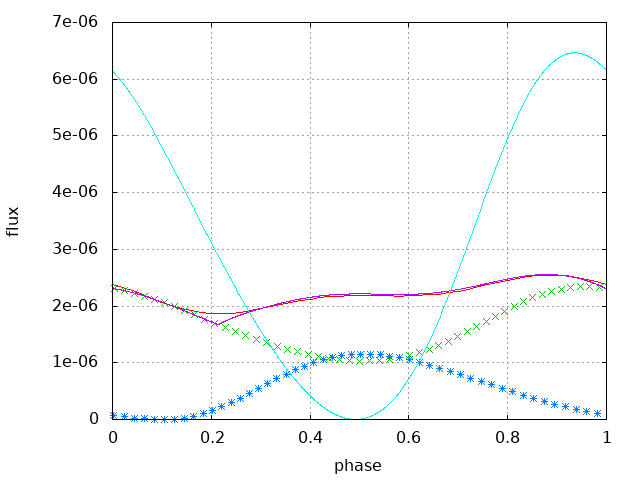} 
\caption{Flux received in the relativistic case, with approximation~(\ref{eq:Beloborodov}), for a line of sight and a magnetic axis that form an angle of $45\degr$ with the rotation axis with in dotted line the emission from each of the polar cap, in red the sum of the two in purple the relativistic case and in blue for the minkowskian case.
}
\label{fig:coupeF45-45RB}
\end{figure}

Compared to flat space-time, general-relativistic effects produce a more homogeneous distribution of the flux with respect to the phase, essentially because of light bending that we already discussed in Section~\ref{sec:image}. The pulse profiles are smeared out. We also note a shift in the phase of the minimum flux received in the minkowskian case compared to the relativistic case, see Fig.~\ref{fig:coupeF45-45RB}. In the GR case, the second pole becomes apparent, which is not the case in the minkowskian case. We kept the information about the absolute intensity in order to see the decrease in flux induced by GR with respect to flat spacetime. 
This shift has nothing to do with a time delay induced by the curvature of the light ray as we can see it on the Fig~\ref{fig:coupeF45-45M} to \ref{fig:coupeF45-45R} where we didn't add the shift in phase due to the photon time of flight.
This shift between the minima of flux received is due to the addition of the two fluxes from each polar cap as in the Schwarzschild metric the hot spots are visible during a greater time for one phase because of light bending explained in Section~\ref{sec:image}.

The time of flight impacts the pulse profiles because several photons can pile up at the time or be smeared in time. Indeed, compared to the profile shown in Fig.~\ref{fig:coupeF45-45R} where the maximum intensity is above $2.5\times10^{-6}$, it is slightly less than $2.5\times10^{-6}$ when Shapiro delay is removed. 

\begin{figure}
\includegraphics[width=\columnwidth]{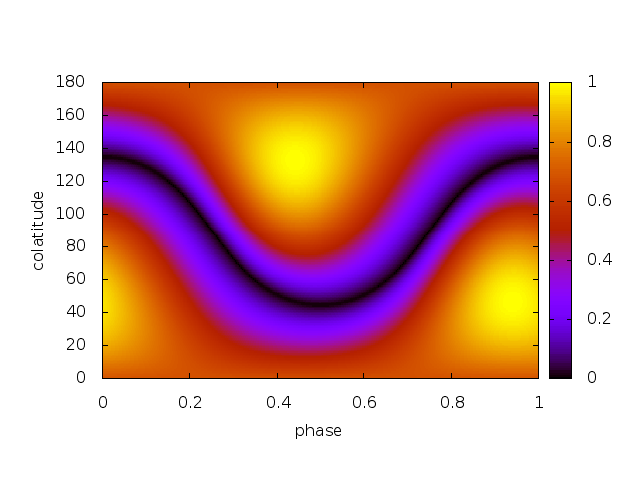}
\caption{Flux received from the two polar gap by a distant observer with an angle of $45\degr$ between the magnetic axis and the rotation axis in the minkowskian case without flight time.}
\label{fig:Flux45_mink_SR}
\end{figure}
\begin{figure} 
\includegraphics[width=\columnwidth]{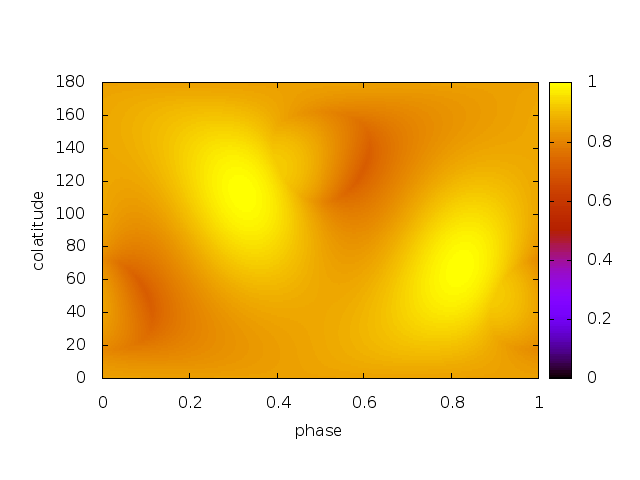} 
\caption{Flux received from the two polar gap by a distant observer with an angle of $45\degr$ between the magnetic axis and the rotation axis in the relativistic case without flight time. }
\label{fig:Flux45_rel_SR}
\end{figure}

A last comparison is performed in Fig.~\ref{fig:Flux45_mink_SR}, where we show sky maps without time of flight effects in minkowskian case and in Fig.~\ref{fig:Flux45_rel_SR} for general relativity. In GR, we notice a change in the phase region where the flux is minimum, around phase $\phi=0.2$ and  phase $\phi=0.6$. Accurate pulse profile modelling therefore requires a careful analysis of the Shapiro delay for realistic investigation of the neutron star surface emission.

In order to increase computational speed or to perform analytical work, an approximate expression is used for light bending as found by \cite{beloborodov_gravitational_2002}. It replaces the integral eq.~(\ref{eq:trajectory_u}) by the simpler expression
\begin{equation}
\label{eq:Beloborodov}
1 - \cos \phi = ( 1 -\cos \theta ) \left( 1 - \frac{\Rs}{R} \right) .
\end{equation}
This expression, although simple, is precise enough for realistic neutron star compactnesses. In Fig.~\ref{fig:coupeF45-45RB}, we compute the flux expected from Beloborodov approximation.

\begin{figure}
\includegraphics[width=\columnwidth]{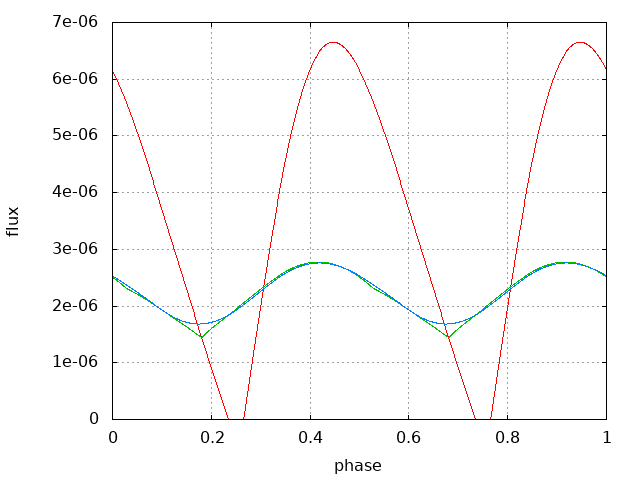} 
\caption{Flux received from the two polar caps for a line of sight and a magnetic axis perpendicular to the rotation axis in a flat space-time (in red), in the relativistic case (in green) and with Beloborodov approximation in blue. }
\label{fig:coupeF90-90}
\end{figure}

The difference between minkowskian and GR is substantial as seen for instance for the orthogonal rotator in the equatorial plane, Fig.~\ref{fig:coupeF90-90}.
It shows the very good agreement between \cite{beloborodov_gravitational_2002} and GR computations.

Next we switch to the impact of GR on the pulsed high energy magnetospheric emission, including light bending and Shapiro delay.

\section{Magnetospheric Emission}
\label{sec:emission}

This section describes the generalization of the work presented in \cite{petri_general-relativistic_2018} by including the Shapiro delay in addition to the light bending and general-relativistic electromagnetic field.

\subsection{Geometry of the magnetic field and Shapiro time delay}
\label{sec:map} 

To simulate the magnetospheric emission of a pulsar, we consider a model where there are gaps in the co-rotating plasma located along the last closed magnetic field lines and over the polar cap \citep{1975ApJ...196...51R, 2008ApJ...680.1378H}.
We trace the magnetic field lines of the neutron star by using generalization of the Hankel's function as presented in \cite{petri_general-relativistic_2018}. As particles are accelerated inside the gaps, we can simulate the emission of photons by curvature radiation, assuming that they are emitted tangentially to the lasts closed magnetic field lines as viewed from the corotating frame. To have an idea of what a distant observer will perceive from this emission, we compute the coordinates of the photon when it impacts on a celestial sphere centred on the neutron star with a radius large enough (for concreteness set to thousand times the light cylinder radius) to minder the influence of the gravitational field so that photons move on straight lines to good accuracy when hitting this sphere. 

We compare the sky maps in two limiting cases of space-time metrics, namely
\begin{itemize}
\item in the minkowskian case, we simply trace the tangent lines to the last closed field lines, 
and then add a phase shift to take into account the photon time of flight, the phase being the longitudinal coordinate on the celestial sphere.
\item in the relativistic case, these maps are obtained by integrating the equations~(\ref{eq:trajectory_u}) and (\ref{eq:temps_u}).
\end{itemize}
The reported differences between the flat space-time and the Schwarzchild metric are due to
\begin{figure}
	\includegraphics[width=\columnwidth]{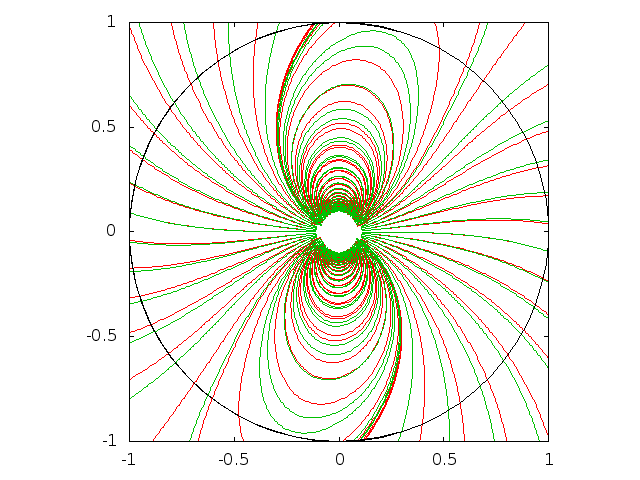}
	\caption{Magnetic field lines of the pulsar in the minkowskian case (in red) and in the relativistic case (in green) in the equatorial plane when the magnetic axis is perpendicular to the rotation axis and with (in black) the light cylinder.} 
	\label{fig:champ_mag}
\end{figure}
\begin{itemize}
\item differences in the geometry of the magnetic field lines between the two cases, as we can see on Fig.~\ref{fig:champ_mag} for the perpendicular rotator.
\item curvature of the photon's trajectory according to general relativity.
\item time delay generated by the curvature of spacetime called the Shapiro delay.
\end{itemize}

\begin{figure}
	\includegraphics[width=\columnwidth]{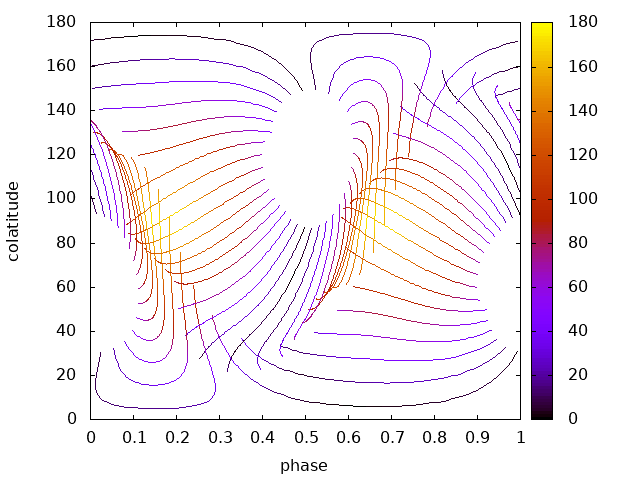}
	\caption{Projection of the minkowskian magnetic field lines without photon bending but with time of flight and aberration effects in minkowskian geometry. The colour scale depicts the angle in degree between the emission direction of the photon at production site and its final direction projected onto the celestial sphere due to aberration and retardation. \label{fig:proj60M-aberr-retard}}
\end{figure}

\begin{figure}
	\includegraphics[width=\columnwidth]{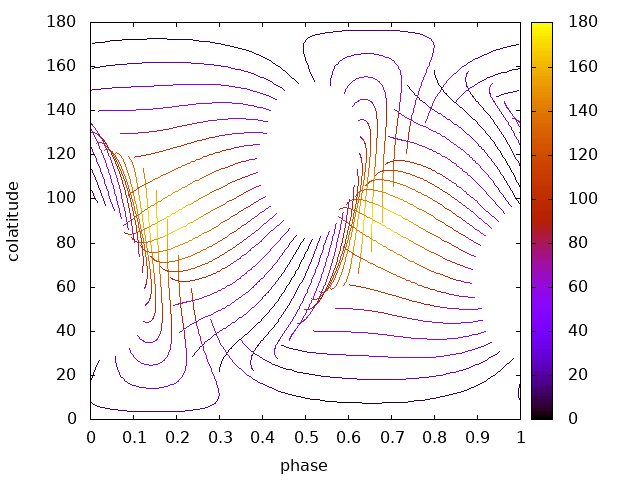}
	\caption{Projection of the GR magnetic field lines with light bending, aberration and Shapiro delay included. The colour scale depicts the angle in degree between the emission direction of the photon at production site and its final direction projected onto the celestial sphere. \label{fig:proj60GR-aberr-retard-courbure}} 
\end{figure}

\begin{figure}
	\includegraphics[width=\columnwidth]{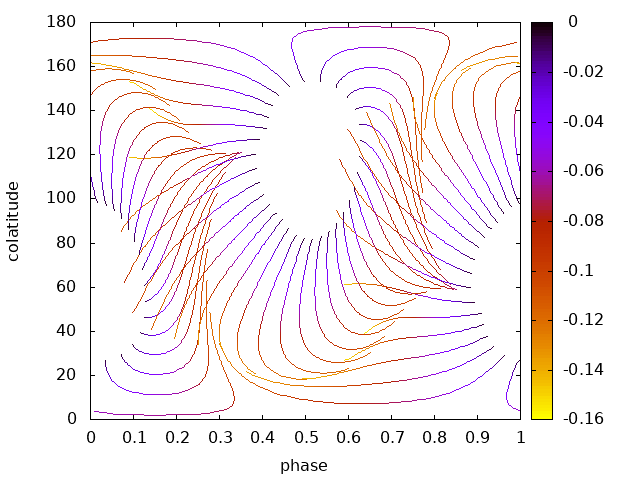}
	\caption{Projection of the GR magnetic field lines with photon bending and Shapiro delay. The colour scale depicts the time delay between the photon and a reference time (that of the photon emitted at the magnetic axis) expressed in fraction of the phase. \label{fig:proj60GR-retard_ref}}
\end{figure}

\begin{figure}
	\includegraphics[width=\columnwidth]{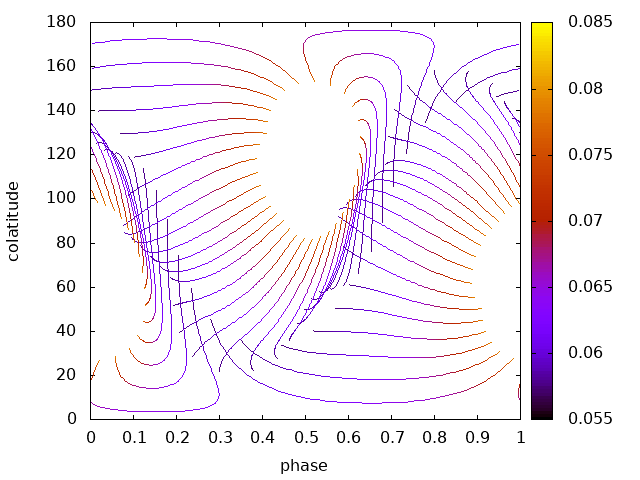}
	\caption{Projection of the GR magnetic field lines with photon bending, Shapiro delay and aberration. The colour scale depicts the difference between the Shapiro time delay and the minkowskian time delay expressed in fraction of the phase.\label{fig:proj60GR-retard_2}}
\end{figure}

In Fig.~\ref{fig:proj60M-aberr-retard} to Fig.~\ref{fig:proj60GR-retard_2} we show the effect of these different factors on the photon impact on the celestial sphere for the special case $\chi=60\degres$. 
Note that for each of these figures, the null phase $\phi=0$ is defined as the date when the observer receives a photon from the magnetic north pole thus around the line of sight $\zeta=60\degres$.

We now detail the merit of aberration, retardation and light bending in the construction of light curves and sky maps. In Fig.~\ref{fig:proj60M-aberr-retard} we show the change in the photon direction angle when aberration and retardation are added in minkowskian geometry. Aberration remains weak as long as photons are produced well within the light-cylinder. This is seen by the colour scale where at the polar caps the deviation is irrelevant. However, when approaching the light-cylinder, the corotation drastically shifts the direction of propagation and retardation effects become important.

Fig.~\ref{fig:proj60GR-aberr-retard-courbure} shows the change in the photon direction from its emission site to infinity when all GR effects are included. The change in angle remains very similar to the minkowskian case. This is due to the fact that photon are emitted almost radially outwards with small angles~$\theta$ is defined by the impact parameter~$b$ in Eq.(\ref{eq:impact}). Indeed when $\theta\ll1$ the light-bending induces by space-time curvature is weak, explaining the good agreement between GR and minkowskian cases. Note however that the shape of the polar cap is slightly modified by GR and becomes larger due to the combined effect of light bending and magnetic field distortion. For an off-centred dipole or more generally non dipolar fields showing a large angle between field lines and radial direction close to the surface, we expect larger discrepancies between minkowskian and GR radiative properties.

Fig.~\ref{fig:proj60GR-retard_ref} shows the time of flight difference between a reference photon taken to be at the magnetic axis and an arbitrary photon emitted locally tangentially to magnetic field lines, normalized to a full period. The advance in time of photon coming from the neighbourhood of the light cylinder is almost 16\%. This number come from the delay of an almost straight propagation of photon from the surface to the light-cylinder, which is given by \cite{petri_unified_2011} 
\begin{equation}\label{key}
\Delta t = \frac{\rlight-R}{2\,\upi\,\rlight} \approx 0.16
\end{equation}
The approximation is valid for slow rotators with~$R\ll \rlight$.

Finally, Fig.\ref{fig:proj60GR-retard_2} shows the error in the photon arrival time when the Shapiro delay is replaced by the minkowskian time of flight approximation. There is a minimum additional time of about 5\% of the period in the vicinity of the light-cylinder up to 8\% of the period near the surface. We conclude that Shapiro delay amounts to 3\% difference in the arrival time of radio pulses with respect to high-energy gamma-ray pulses.

\subsection{High-energy sky maps}
\label{sec:high_energy_map}

\begin{figure}
\includegraphics[width=\columnwidth]{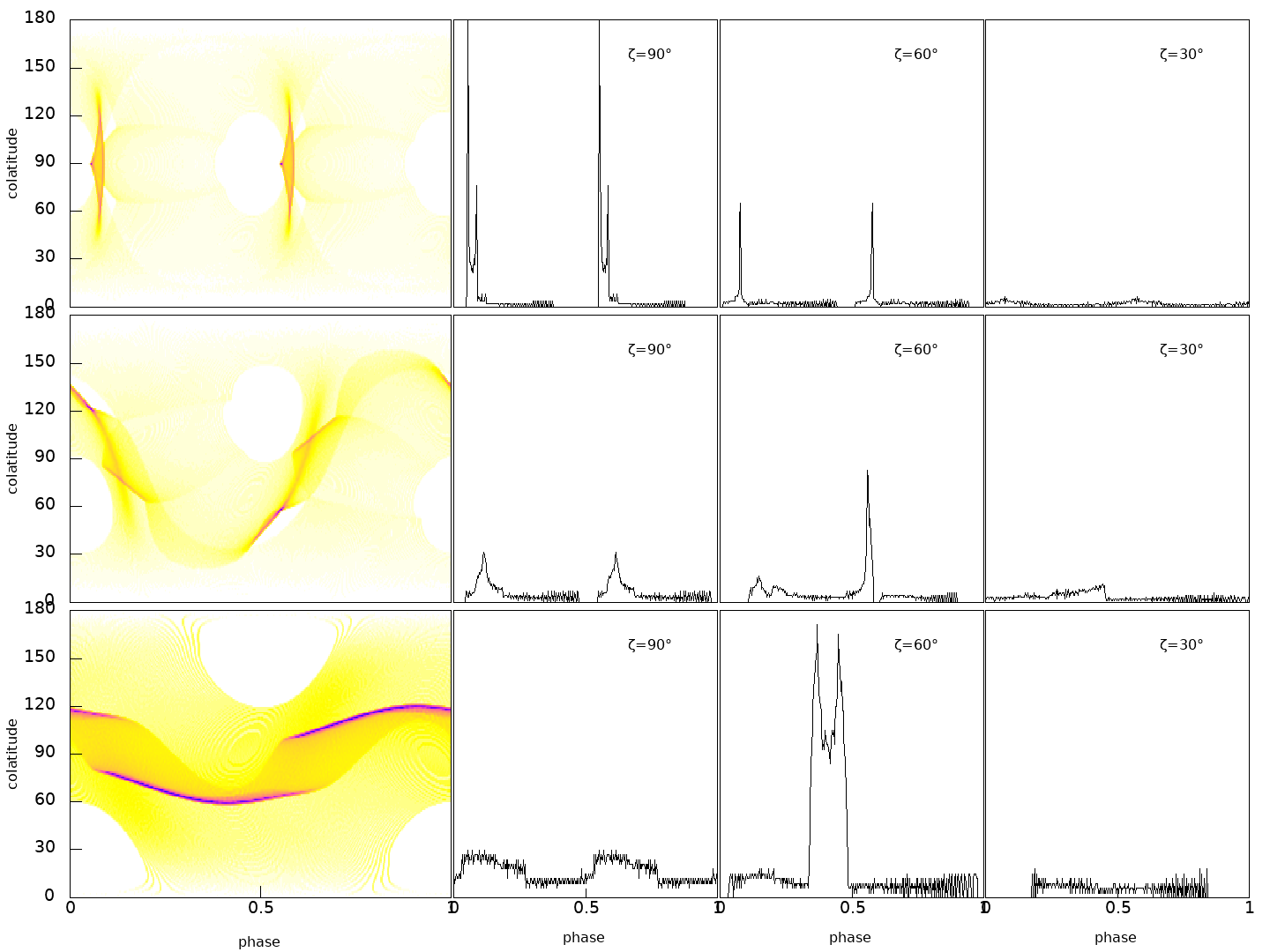}
\caption{Emission maps for different obliquities~$\chi$ (from top to bottom : $90^{\circ}$, $60^{\circ}$ and $30^\circ$) for the minkowskian case with light curves for some several values of the inclination angle~$\zeta$ (from left to right : $90^{\circ}$, $60^{\circ}$ and $30^\circ$).}
\label{fig:emission_HE_min}
\end{figure}

\begin{figure}
\includegraphics[width=\columnwidth]{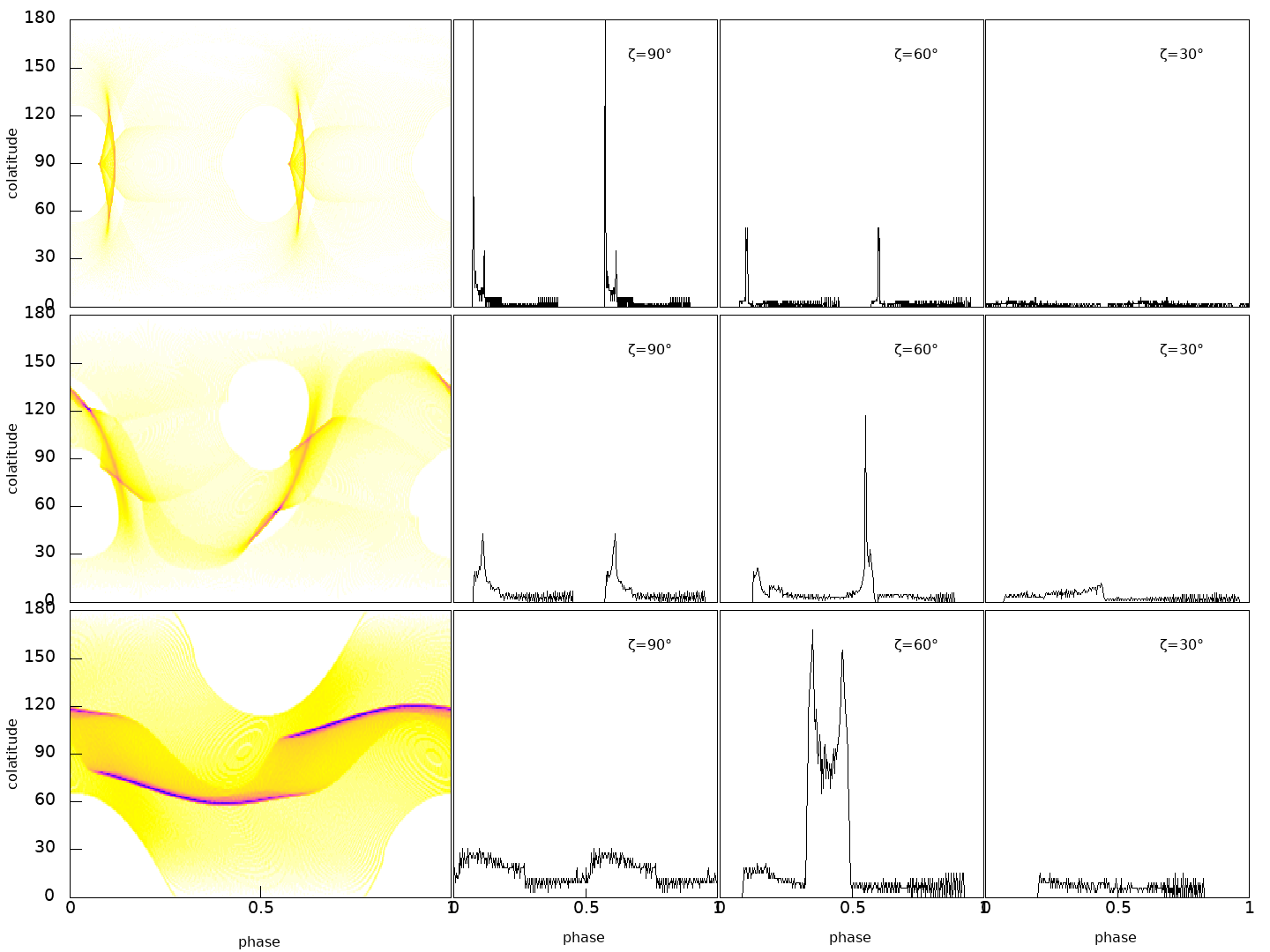}
\caption{Emission maps for different obliquities~$\chi$ (from top to bottom : $90^{\circ}$, $60^{\circ}$ and $30^\circ$) for the relativistic case with light curves for some several values of the inclination angle~$\zeta$ (from left to right : $90^{\circ}$, $60^{\circ}$ and $30^\circ$).}
\label{fig:emission_HE_rel}
\end{figure}

Sky maps are a good mean to synthesize the impact of the viewing angle onto the pulse profile. Fig.~\ref{fig:emission_HE_min} and Fig.~\ref{fig:emission_HE_rel} show typical examples comparing flat to curved spacetime respectively with obliquities~$\chi\in\{30\degr,60\degr,90\degr\}$, with a relevant sample of light-curves assuming inclination of the light of sight of $\zeta\in\{30\degr,60\degr,90\degr\}$. These maps are drawn following the same procedure as for the maps of photon impacts on the celestial sphere in Section~\ref{sec:map} but with a colour code describing the intensity (actually the number of photons) of perceived radiation and by adding the aberration effect seen in Section~\ref{sec:aberration} (maximum intensity in red/black and low intensity in yellow/white). The sky maps are synchronized with the reception of the radio photon emitting at the magnetic axis (taking to be phase zero in the plot). Comparing plots from Fig.~\ref{fig:emission_HE_min} and Fig.~\ref{fig:emission_HE_rel}, we conclude that GR tends to smear the light curves and to decrease the peak intensity levels. Indeed, as summarized in table~\ref{tab:intensite_gamma}, the decrease is significant for the perpendicular rotator whereas it almost vanishes for close to aligned rotators. This is partly due to light bending, spreading the photons on a broader solid angle, and because of additional delays induced by the Shapiro delay. The individual pulses are very sharp because we assumed emission only from the last closed magnetic surface. More realistically, we would expected a widening of the pulse profiles associated the thickness of this surface as done artificially for instance by \cite{dyks_relativistic_2004} and \cite{bai_uncertainties_2010}. There is indeed no physical constraint to estimate the size of this layer except maybe by fitting to gamma-ray light-curves of a sample of Fermi pulsars \citep{abdo_second_2013}. In all cases, the pulses become narrower and extremely sharp for perpendicular rotators.

\begin{table}
	\centering
	\caption{Maximum intensity in high energy for various angles between the magnetic axis and the rotation axis.}
	\label{tab:intensite_gamma}
	\begin{tabular}{lcccr} 
		\hline
		obliquity~$\chi$ & $90^{\circ}$ & $60^{\circ}$ & $45^{\circ}$ & $30^{\circ}$\\
		\hline		\hline
		minkowskian & 195 & 154 & 64 & 60\\
		relativistic & 87 & 75 & 65 & 58\\
		\hline
	\end{tabular}
\end{table}

\subsection{Radio sky maps}
\label{sec:radio_maps}

\begin{figure}
\includegraphics[width=\columnwidth]{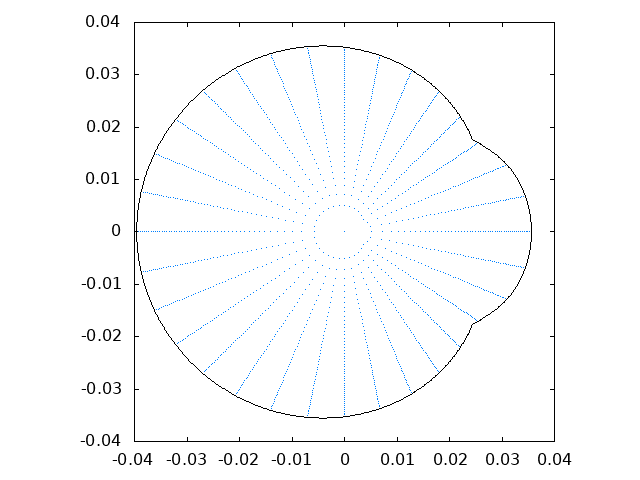} 
\caption{Example of point distribution (in blue) inside the polar cap (in black). The magnetic axis, located at the origin, is perpendicular to the rotation axis.}
\label{fig:polar_cap}
\end{figure}

Pulsar radio emission originates from above the polar caps, well within the light-cylinder, where the magnetic field lines are almost rectilinear and dipolar. To simulate this particular emission, we spread emission points over the entire stellar surface in the area delimited by the last closed field lines, those points are placed between the magnetic north and south poles and each intersection between the star's surface and one of the last closed field lines. The spacing between those points is designed to have a homogeneous density of emission points on the polar cap. Expressed in the frame oriented along the magnetic moment, fixing a value of the longitude, the latitude $\theta$ of each of these points~$i\in[0..N]$ is determined by the formula
\begin{equation}
\theta - \theta_{\rm pc}= \theta_{\rm mp} \, \left( 1 - \sqrt{\frac{i}{N}} \right) .
\end{equation}
with $\theta_{\rm mp}$ the latitude of the magnetic pole ($0^{\circ}$ and $180^{\circ}$ for each pole in the frame oriented along the magnetic moment), $\theta_{\rm pc}$ is the latitude of the point where the last closed field line crosses the star surface, $N$ is the number of points we desire between the pole and the rim of the polar cap. The square root dependence is introduced to keep a constant surface density of sampling points, avoiding an artificial concentration around the magnetic poles. We then shoot single photons from each of these points and compute their impact on the celestial sphere taking into account all propagation effects. Fig.~\ref{fig:polar_cap} shows an example of this sampling for $N=50$ points between the centre and the rim of one polar cap.

\begin{figure}
\includegraphics[width=\columnwidth]{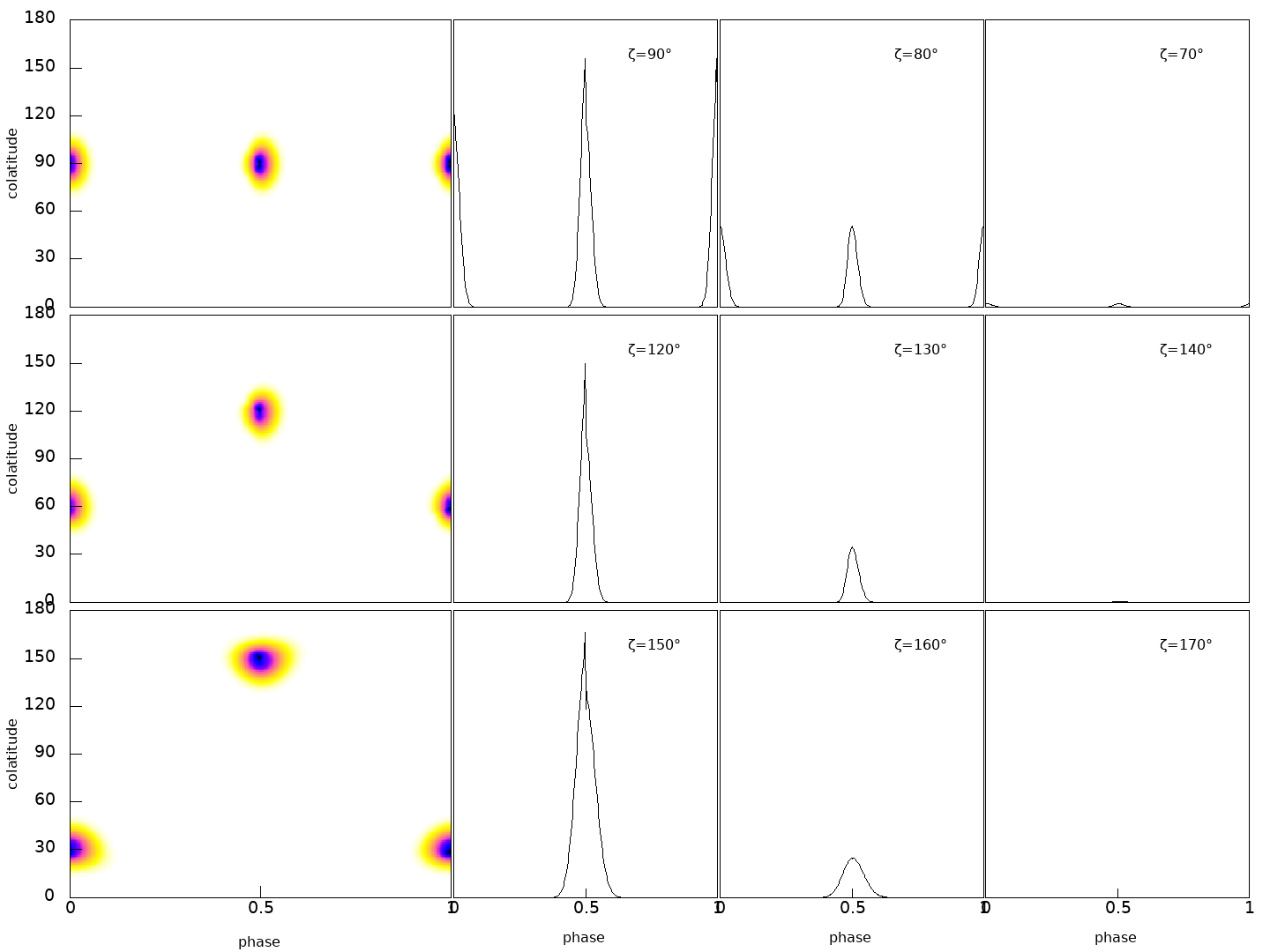}
\caption{Radio emission for different angles $\chi$ of the magnetic axis (from top to bottom : $90^{\circ}$, $60^{\circ}$ and $30^\circ$) for the minkowskian case with light curves for some several values of the inclination angle $\zeta$.}
\label{fig:emission_radio_mink}
\end{figure}

\begin{figure}
\includegraphics[width=\columnwidth]{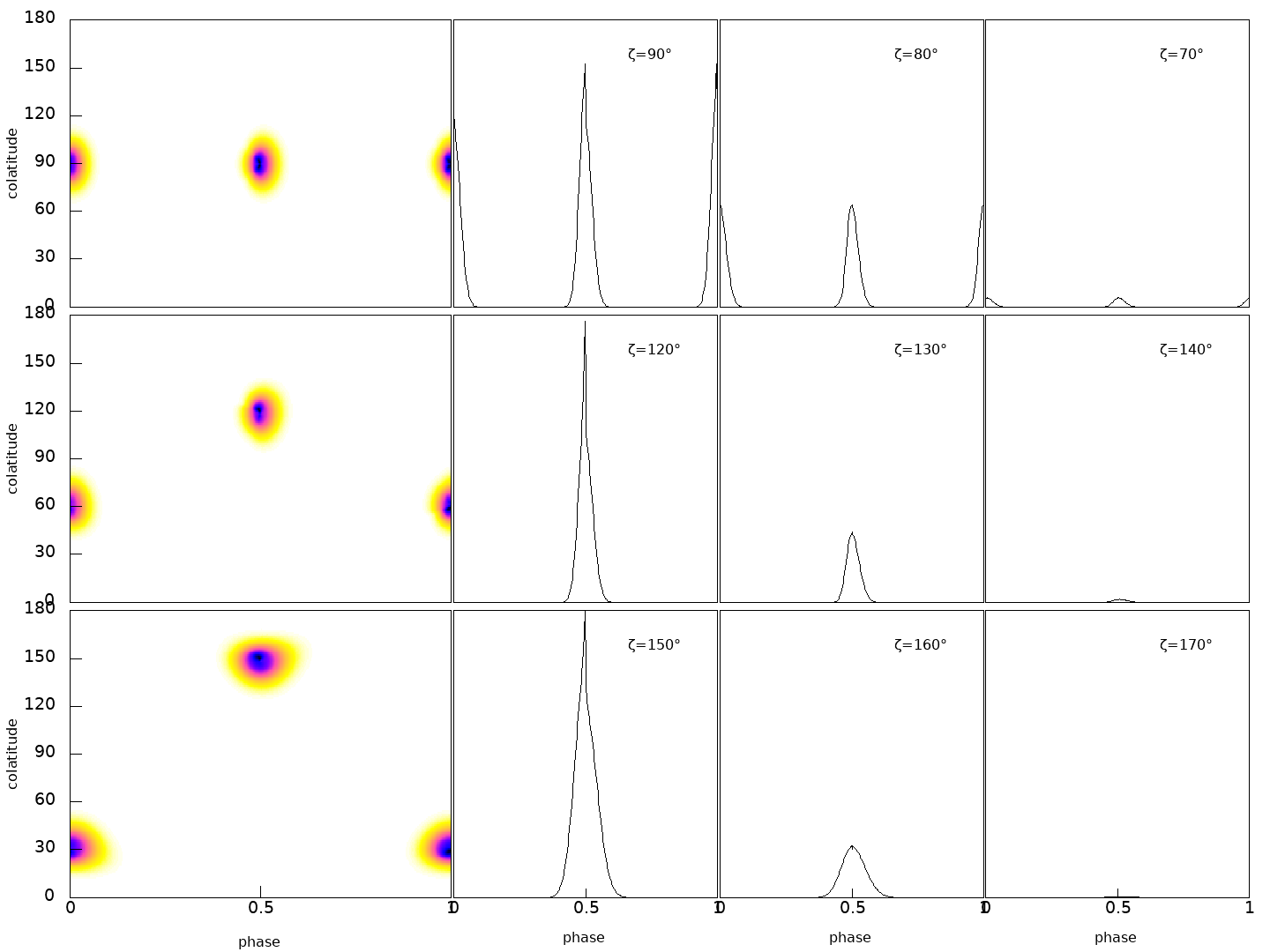}
\caption{Radio emission for different angles $\chi$ of the magnetic axis (from top to bottom : $90^{\circ}$, $60^{\circ}$ and $30^\circ$) for the relativistic case with light curves for some several values of the inclination angle $\zeta$.}
\label{fig:emission_radio_rel}
\end{figure}

To have realistic radio pulse profiles, similar to those observed, we attributed to every received photon a weight depending on its initial position to simulate sky maps having Gaussian radio intensity profiles such that the weight is given by
\begin{equation}
w(\theta) = e^{\frac{-\theta^{2}  / \theta_{\rm pc}^{2}}{\sigma_{\rm pc}^{2}}}
\end{equation} 
with the width of the Gaussian controlled by~$\sigma_{\rm pc}$ chosen equal to $\frac{1}{\sqrt{10}}$. With these parameters, we obtain the emission maps shown in Fig.~\ref{fig:emission_radio_mink} for Minkowski spacetime and Fig.~\ref{fig:emission_radio_rel} for Schwarzschild spacetime.
Here again, GR smears the pulse profile and minders the maximum intensity as reported quantitatively in Table~\ref{tab:intensite_radio}.

\begin{figure}
\includegraphics[width=\columnwidth]{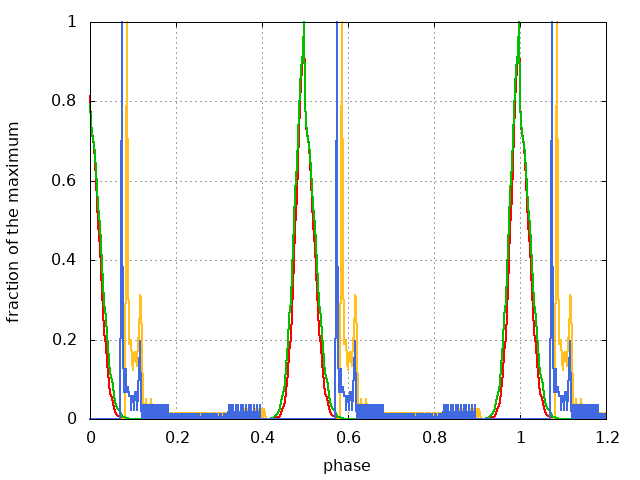} 
\caption{Radio and high-energy light-curves for $\chi=90\degr$ and $\zeta=90\degr$. Radio emission in red for a flat space-time and in green for the relativistic case, high-energy emission in orange for a flat space-time and in blue for the relativistic case. }
\label{fig:total90-90}
\end{figure}

\begin{figure}
\includegraphics[width=\columnwidth]{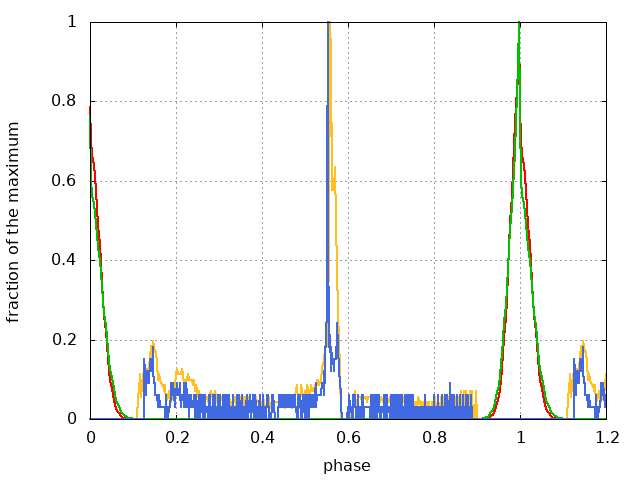}
\caption{Radio and high-energy light-curves for $\chi=60\degr$ and $\zeta=60\degr$. Radio emission in red for a flat space-time and in green for the relativistic case, high-energy emission in orange for a flat space-time and in blue for the relativistic case.} 
\label{fig:total60-60}
\end{figure}

\begin{figure}
\includegraphics[width=\columnwidth]{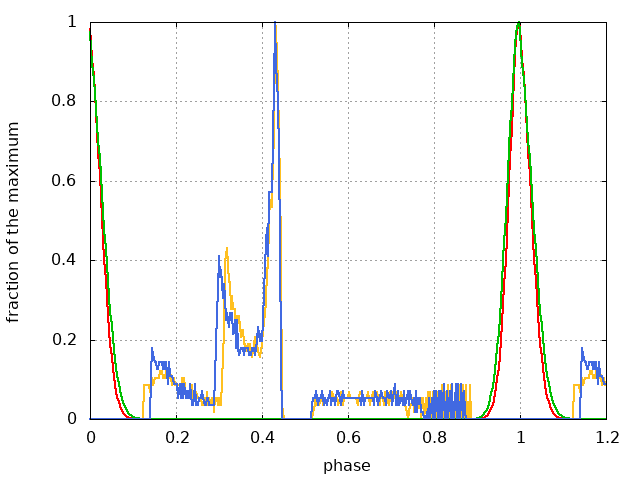} 
\caption{Radio and high-energy light-curves for $\zeta=50^{\circ}$ and $\chi=45^{\circ}$. Radio emission in red for a flat space-time and in green for the relativistic case, high-energy emission in orange for a flat space-time and in blue for the relativistic case.} 
\label{fig:total45-50}
\end{figure}

\begin{figure}
\includegraphics[width=\columnwidth]{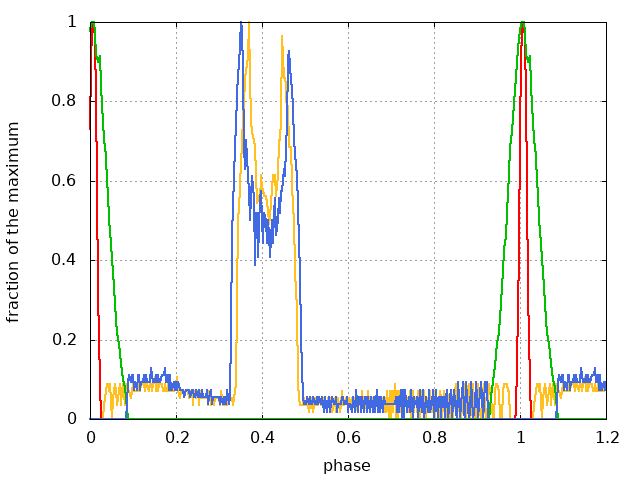} 
\caption{Radio and high-energy light-curves for $\zeta=60^{\circ}$ and $\chi=30^{\circ}$. Radio emission in red for a flat space-time and in green for the relativistic case, high-energy emission in orange for a flat space-time and in blue for the relativistic case.}
\label{fig:total30-60}
\end{figure}

\begin{table}
	\centering
	\caption{Maximum intensity in the radio band for different angles between the magnetic axis and the rotation axis.}
	\label{tab:intensite_radio}
	\begin{tabular}{lcccr} 
		\hline
		obliquity~$\chi$ & $90^{\circ}$ & $60^{\circ}$ & $45^{\circ}$ & $30^{\circ}$\\
		\hline		\hline
		minkowskian & 4108 & 4376 & 4007 & 2510 \\
		relativistic & 3194 & 3381 & 3138 & 1986 \\
		\hline
	\end{tabular}
\end{table}

\subsection{Multi-wavelength light curves}
\label{sec:lightcurves}

Finally, in order to better compare the full impact of GR on pulsar emission, we plot several representative multi-wavelength light-curves extracted from the sky maps presented in Section~\ref{sec:high_energy_map} and Sec.~\ref{sec:radio_maps} which are in fact light curves of the pulsar for different angles between the line of sight and the rotation axis. In Fig.~\ref{fig:total90-90} to \ref{fig:total30-60}, we plot these light curves for one value of the line of sight inclination angle for both radio and high-energy emission (in arbitrary units normalizing the peak intensity for a better visibility of the pulses). We see a difference in the pulse shape, both in radio and high-energy emission, depending on the viewing geometry. For instance, when the line of sight is grazing the polar cap rim, because the Schwarzschild metric broadens the pulses, the radio profiles are very different in this case as it is noticeable in Fig.~\ref{fig:total30-60}.

\begin{figure}
	\centering
	\includegraphics[width=0.9\linewidth]{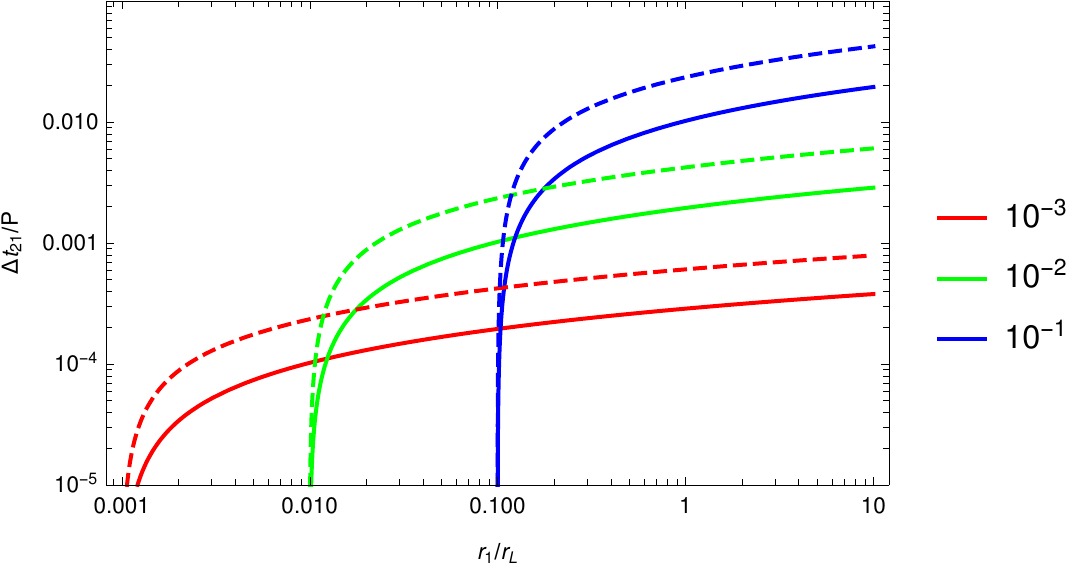}
	\caption{Shapiro time delay~$\Delta t_{21}/P$ for several spin parameters $a=R/\rlight=\{10^{-3}, 10^{-2}, 10^{-1}\}$ in red, green and blue respectively, and two compactnesses $K=0.25$ in solid line and $K=0.5$ in dashed line, assuming one photon coming from the surface.}
	\label{fig:shapirodelay}
\end{figure}

The additional time lag between radio and gamma-ray pulses can be estimated by the following simple argument. Consider photons produces at two emission heights labelled respectively by~$r_1$ and $r_2$. For photons propagating in the radial direction, integration in the Schwarzschild metric leads to a time lag~$\Delta t_{21}$ between pulse~2 and pulse~1 normalized to the period~$P$ such that such that
\begin{equation}\label{key}
\frac{\Delta t_{21}}{P} = \frac{r_1-r_2}{2\,\upi\,\rlight} + \frac{\Rs}{2\,\upi\,\rlight} \, \ln \left( \frac{r_1-\Rs}{r_2-\Rs}\right) .
\end{equation}
This lag is independent of the distance to the observer. The first term on the right hand side corresponds to the time of flight in flat space-time whereas the second term on the right hand side is due to the space-time curvature and identified as the Shapiro delay. This delay is shown in fig.~\ref{fig:shapirodelay} for several spin parameters defined by $a=R/\rlight$ and two compactnesses $K=0.25$ in solid line and $K=0.5$ in dashed line. We assume that photon number~2 is coming from the surface $r_2=R$ and vary the location of the first photon~$r_1$.
As expected, it is highest for fast spinning and compact neutron stars, reaching additional delay of several percent with respect to flat space-time. Interestingly, in principle the Shapiro increases logarithmically with distance without bounds. If the high-energy photons are coming from well outside the light-cylinder, as claimed for the striped wind model, this delay can be increased by about a factor two at $10\rlight$.

\section{Conclusion}
\label{sec:conclusions}

In this paper, we have numerically simulated the effects of the gravitational field of a neutron star on its image, its polar thermal flux and its magnetospheric emission by curvature radiation.
We demonstrated that the gravitational field of a neutron star affects its magnetospheric emission as observed by a distant observer, according to General Relativity although it is important to notice that this effect can be pretty subtle and be only a shift in phase that will not be perceptible during a normal observation. Thermal emission from the polar gap seems to be a more sensible indicator of the presence of relativistic effects in the gravitational field of the neutron star as it seems to affect more significantly the light curve shapes. 

In future developments of our model, we plan to focus on the effect of the gravitational field on the polarization of emitted photons, looking also also for dragging phenomenon, even by using Kerr metric instead of the Schwarzschild's one, gravitational redshift and also other mechanism of magnetospheric emission as synchrotron radiation or inverse Compton effect, for a more complete approach of how relativistic effects affect the properties of pulsar radiation.

Whence the model is fully complete, including self-consistently all GR effect, we will apply our model to some pulsars simultaneously detected in radio and in the high-energy MeV/GeV band as reported by the second Fermi catalogue \cite{abdo_second_2013}.

\section*{Acknowledgements}
This work has been partly supported by CEFIPRA grant IFC/F5904-B/2018.

\appendix

\section{Trajectory of a photon in three dimension}
\label{app:A}

As the trajectory of a photon in the Schwarzschild metric is always contained in a plane, we can trace its path in space by first tracing it into a two-dimensional plane and then use the Euler rotation matrix to switch back to full three dimensions.

Introducing the three Euler angle as
\begin{itemize}
	\item the precession~$\alpha$
	\item the nutation~$\beta$
	\item the proper rotation~$\gamma$
\end{itemize}
\begin{small}
\begin{equation}
\begin{pmatrix}
\cos \alpha \cos \gamma - \sin \alpha \cos \beta \sin \gamma &-\cos \alpha \sin \gamma -\sin \alpha \cos \beta \cos \gamma & \sin \alpha \sin \beta\\
\sin \alpha \cos \gamma + \cos \alpha \cos \beta \sin \gamma &-\sin \alpha \sin \gamma +\cos \alpha \cos \beta \cos \gamma & -\cos \alpha \sin \beta\\
\sin \beta \sin \gamma & \sin \beta \cos \gamma & \cos \beta
\end{pmatrix}
\label{eq:Euler}
\end{equation}
\end{small}

First we assume a new frame where a point is located by its coordinate $x'$, $y'$ and $z'$ and where the emission point of a photon and its initial direction of propagation are contained in the plane $z'=0$. To obtain the coordinate of a point, as the emission point, in the new frame from the ones in the initial frame ($x$, $y$ and $z$), we will use the relation deduced from the Euler rotation matrix~(\ref{eq:Euler}) such that
\begin{equation}
\begin{pmatrix}
x'\\ y'\\ z'\\
\end{pmatrix}
= 
\begin{pmatrix}
\cos A  & \sin A & 0\\
-\sin A \cos B & \cos A \cos B & \sin B\\
\sin B\sin A & -\sin A \cos A & \cos B
\end{pmatrix}
\begin{pmatrix}
x\\
y\\
z\\
\end{pmatrix}
\end{equation}
By calling $p$ the intersection point of the plane $z=0$ and the photon's initial direction of propagation
\begin{itemize}
\item $A$ is the angle between the $x$ axis and a line passing through the origin and $p$,
\item $B$ is the angle between the plane $z=0$ and the photon's initial direction of propagation.
\end{itemize}
After tracing the trajectory of the photon in the plane $z'=0$,  we can deduce the coordinates in space, in the initial frame, of any point of the trajectory by using the reciprocal transformation given by 
\begin{equation}
\begin{pmatrix}
x\\ y\\ z\\
\end{pmatrix}
= 
\begin{pmatrix}
\cos A & -\sin A \cos B & \sin A \sin B\\
\sin A & \cos A \cos B & -\cos A \sin B\\
0 & \sin B & \cos B
\end{pmatrix}
\begin{pmatrix}
x'\\
y'\\
z'\\
\end{pmatrix}
\end{equation}
This trick unfortunately only works for spherically symmetric spacetimes. For rotating metrics, we would have to perform directly full three-dimensional integrations.

\bsp	
\label{lastpage}
\end{document}